\documentclass[letterpaper,twocolumn,10pt]{article}
\usepackage{usenix}
\usepackage[available]{usenixbadges}
\usepackage{tikz}
\usepackage{bbm}
\usepackage{amsmath}
\usepackage{subcaption}
\usepackage{booktabs}
\usepackage{color}
\usepackage{tabularx}
\usepackage{enumitem} 

\begin{document}

\date{}

\title{\Large \bf COGNITION: From Evaluation to Defense against Multimodal LLM CAPTCHA Solvers}

\author{
{\rm Junyu Wang}$^{1}$,
{\rm Changjia Zhu}$^{2}$,
{\rm Yuanbo Zhou}$^{1}$,\\
{\rm Lingyao Li}$^{2}$,
{\rm Xu He}$^{3}$,
{\rm Mingkui Wei}$^{4}$,
{\rm Junjie Xiong}$^{1}$\thanks{Corresponding author.}\\[0.5ex]
$^{1}$Missouri University of Science and Technology;\\
$^{2}$University of South Florida;
$^{3}$Visa Inc.;
$^{4}$George Mason University
} 

\maketitle

\begin{abstract}
This paper studies how multimodal large language models (MLLMs) undermine the security guarantees of visual CAPTCHA. We identify the attack surface where an adversary can cheaply automate CAPTCHA solving using off-the-shelf models. We evaluate 7 representative MLLMs on 18 real-world CAPTCHA task types, measuring single-shot accuracy, success under limited retries, end-to-end latency, and per-solve cost. We further validate our findings through a supplemental external dataset and an adaptive-attacker setting with session memory, while also analyzing the impact of task-specific prompt engineering and few-shot demonstrations on solver effectiveness. We reveal that MLLMs can reliably solve recognition-oriented and low-interaction CAPTCHA tasks at human-like cost and latency, whereas tasks requiring fine-grained localization, multi-step spatial reasoning, or cross-frame consistency remain significantly harder for current models. By examining the reasoning traces of such MLLMs, we investigate the underlying mechanisms of why models succeed/fail on specific CAPTCHA puzzles and use these insights to derive defense-oriented guidelines for selecting and strengthening CAPTCHA tasks. To validate these principles, we present a proof-of-concept by hardening a vulnerable CAPTCHA type using our guidelines. We demonstrate that incorporating fine-grained localization and implicit counting reduces the success rate of state-of-the-art MLLMs from over 95\% to 0\%, confirming that structural changes can effectively mitigate the threat. We conclude by emphasizing the urgent need for CAPTCHA redesign as MLLM capabilities increasingly threaten existing defenses.
Code Availability\footnote{https://doi.org/10.5281/zenodo.20406852}.
\end{abstract}

\section{Introduction}

Today, visual CAPTCHA \textit{(Completely Automated Public Turing test to tell Computers and Humans Apart)} challenges are an essential security mechanism in the modern web ecosystem. CAPTCHA has been integrated into more than 10-million commercial websites worldwide, reflecting its role as a standard security layer in online services~\cite{builtwith_recaptcha_usage}. Online services rely on CAPTCHAs to distinguish human users from automated scripts, thereby protecting registration flows, content submission portals, and valuable resources from large-scale abuse. Their ubiquity reflects their continued importance in securing everyday online interactions.

However, the advancement of multimodal large language models (MLLMs) poses a serious threat to this web security mechanism. Unlike prior attack methods that require task-specific training data and architectures~\cite{Sivakorn2016, laperdrix2020browser, 10646606}, MLLMs can jointly interpret images and natural language instructions through a unified interface, precisely the capability that modern CAPTCHAs assume only humans possess. Recent work has demonstrated that MLLMs can serve as general-purpose CAPTCHA solvers~\cite{teoh2025captchas, deng2024oedipus}, building on earlier efforts that evaluated multimodal models under fixed templates or curated synthetic datasets~\cite{nopecha_developers_2025, theyka2025turnstilesolver, dessant2020buster}. Meanwhile, commercial CAPTCHA-solving services, often referred to as CAPTCHA farms~\cite{verifiedvisitors_captcha_farms}, have begun combining MLLM automation with human labor to improve throughput~\cite{twocaptcha_service, captchacoder_service, decaptcher_service}. If off-the-shelf MLLMs can reliably bypass CAPTCHAs at low cost and high speed, the consequences for web security are severe: attackers could automate account creation, credential stuffing, content spam, and resource scraping at unprecedented scale.

This security concern motivates our study. Rather than merely confirming that MLLMs can solve CAPTCHAs, we ask a more critical question: \emph{which CAPTCHA designs can still raise the bar against MLLM-based solvers?} Existing studies mostly report accuracy on narrow benchmarks while overlooking solving time, retry limits, and monetary cost that determine whether an attack is viable at scale~\cite{zhang2020robust, Plesner_2024}, and they rarely analyze why models succeed on some tasks but systematically fail on others~\cite{captchaworld25}. Motivated by these gaps, we consider the following research questions:

\begin{itemize}
\item RQ1: How well can MLLMs solve diverse visual CAPTCHA tasks under practical constraints such as limited time and retries?
\item RQ2: How do prompting strategies, such as direct prompting, optimized instructions, and few-shot demonstrations, affect solver performance across different CAPTCHA types?
\item RQ3: What reasoning behaviors do MLLMs exhibit during successful and failed attempts, and what do these behaviors reveal about the challenges that CAPTCHAs pose to automated solvers?
\item RQ4: What strategies should web service providers adopt to deploy CAPTCHA schemes that remain effective against increasingly capable MLLMs?
\end{itemize}


To answer these questions, we evaluate seven representative MLLMs on 18 real-world visual CAPTCHA types under multiple prompting strategies. Our results reveal a pronounced hardness gap across task types. Recognition-oriented and low-interaction CAPTCHAs, such as Path\_Finder, Select\_Animal, and Image\_Recognition, are already solvable by MLLMs through simulated human reactions within realistic retry and time budgets. In contrast, tasks including Click\_Order, Place\_Dot, Pick\_Area, Dice\_Count, and Patch\_Select, remain consistently challenging: even with strong models and carefully designed prompts, they exhibit low success rates and substantially higher per-solve time and cost. We further study a black-box attack setting in which an adversary repeatedly sends CAPTCHA images and instructions to an off-the-shelf MLLM API, treating each model call as one attempt and stopping until the challenge is solved or a fixed retry cap is reached. Under this setting, most recognition-style CAPTCHAs can be bypassed easily, raising questions about the reliability of current human-verification systems. {In addition, to confirm above findings generalize beyond the main benchmark and remain robust against stronger attacker strategies, we further evaluate a supplemental external dataset and an adaptive session-memory attack setting.}


Importantly, through analysis of reasoning traces from MLLMs, we uncover that the observed robustness of these harder CAPTCHA types is neither accidental nor an artifact of poor prompting. Instead, it reflects deeper structural limitations of current MLLMs. We find that tasks such as Click\_Order, Place\_Dot, Pick\_Area, Dice\_Count, and Patch\_Select share a set of structural properties that consistently stress these models. Specifically, they require fine-grained spatial grounding in continuous image space, precise binding between objects and their locations, and explicit counting or aggregation under visual clutter. While MLLMs often exhibit correct high-level reasoning in natural language, they frequently fail to translate this reasoning into exact coordinates, stable multi-object bindings, or accurate counts. These failure modes persist even under optimized prompts, retries, and few-shot guidance. {Besides identifying which CAPTCHA types remain hard for MLLMs, we analyze the causes of these structural hardness patterns, distill the insights into a concrete defense methodology, and validate it by strengthening a vulnerable CAPTCHA design.}

Building on these observations, our key insight is that CAPTCHA robustness against MLLMs is governed less by semantic recognition difficulty and more by grounding precision and compositional consistency. This insight motivates a defense strategy that shifts CAPTCHA design away from discrete recognition or classification and toward challenges that tightly couple perception with fine-grained localization, ordering, and counting. As a proof of concept, we apply this guideline to structurally redesign an existing MLLM-solvable CAPTCHA task, Select\_Animal, and demonstrate that the success rate of state-of-the-art MLLMs is substantially reduced, decreasing from over 95\% to 0\%.

We summarize our contributions as follows:
\begin{enumerate}[leftmargin=1em, itemsep=0pt, topsep=1pt, parsep=1pt, partopsep=1pt]
\item We formulate a realistic black-box threat model for MLLM-based CAPTCHA solving and propose an evaluation framework that jointly considers accuracy, finite-retry behavior, end-to-end latency, and token-based cost.
\item We empirically evaluate seven MLLMs on 18 real-world CAPTCHA task types, revealing a pronounced and stable hardness gap and identifying which task types are already broken, nearly broken, or still robust. {We further strength this evidence using an external evaluation set and an adaptive session-memory test that examines whether prior feedback enables a stronger attacker to weaken hard task types.}
\item We analyze MLLM reasoning traces on the hardest task types, extract the structural factors underlying their robustness, and {translate these insights into a concrete defense methodology. We validate this methodology through stronger CAPTCHA design transformations.}
\end{enumerate}

\section{Related Work}
\label{sec:related-work}

\textbf{\textit{CAPTCHA Solvers:}} in response to each other's advancement~\cite{ousat2024matter,searles2023modern}. Traditional text- and image-based CAPTCHAs are weakened by deep-learning methods employing convolutional models and segmentation-based techniques~\cite{sivakorn2016robot,tian2020generic,shi2020text}. More advanced architectures, including generative adversarial network-based solvers and transformer-based recognizers, further improved solvers robustness against noisy and diverse schemes~\cite{ye2018yet,zhao2023geesolver}. Besides visual recognition, reinforcement-learning approaches showed that even behavior-based CAPTCHAs can be bypassed by agents learning human-like interaction pattern~\cite{akrout2019hacking, tsingenopoulos2022captcha}. However, the emergence of reasoning-based CAPTCHAs introduces semantics, interactive, and multi-modal designs, necessitating solvers with advanced logical-reasoning capabilities alongside strong image-comprehension skills~\cite{gao2021research,wu2025mca,captchaworld25}.

\begin{figure*}[htbp]
  \centering
  \includegraphics[width=0.74\textwidth]{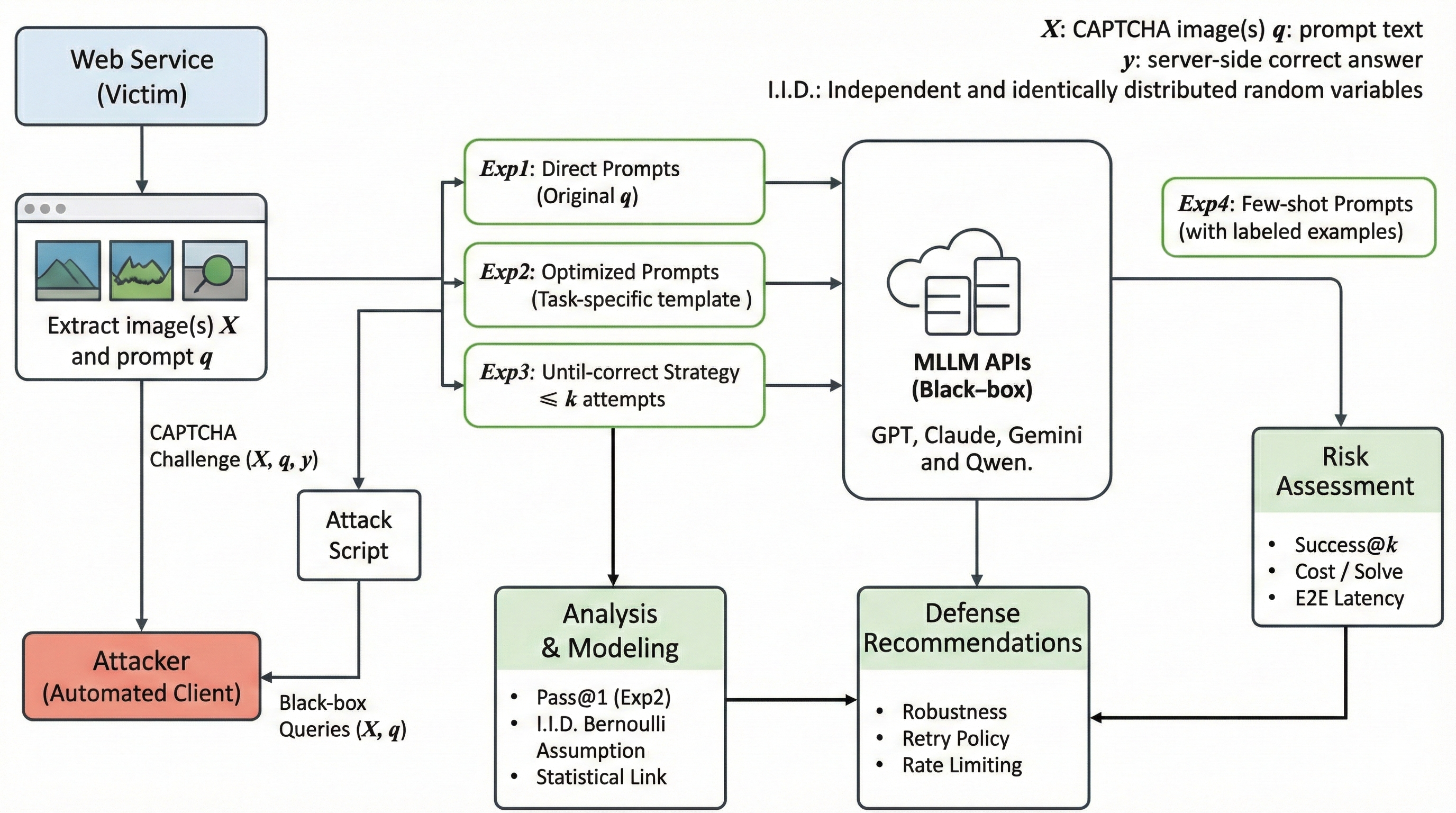}
  \caption{CAPTCHA robustness evaluation framework against MLLMs.}
  \label{fig:framework}
\end{figure*}

\textbf{\textit{MLLM in CAPTCHA:}} With the advent of MLLMs and vision-language models (VLMs), recent work has begun to treat CAPTCHA solving as general interactive reasoning tasks. Teoh et al. propose Halligan, a generalized visual CAPTCHA solver built around an agentic VLM that plans browser actions and achieves high success rates across widely deployed CAPTCHA schemes~\cite{teoh2025captchas}. Deng et al. presents Oedipus, which casts CAPTCHA solving as LLM-guided multi-step reasoning over challenge instructions and candidate visual cues, combining chain-of-thought prompting with perception modules~\cite{deng2024oedipus}. In parallel, Ding et al. design IllusionCAPTCHA which introduces visual illusions that cause state-of-the-art VLM and LLM-based agents to hover near random-guess performance, which illustrates that the carefully crafted perception traps can systematically mislead LLM/VLM CAPTCHA solvers~\cite{ding2025illusioncaptcha}.

Building on this trend, subsequent research has shifted from proposing individual solvers to systematically benchmarking multimodal agents~\cite{captchaworld25, wu2025mca}, analyzing their failure modes~\cite{song2025reasoning}, and designing LLM-aware CAPTCHA defenses~\cite{li2025mi}. However, these efforts largely focus on accuracy and query statistics while overlooking the practical time budgets for CAPTCHA solving, MLLMs deep reasoning (i.e., thinking) capability, and deep analysis on why these models would fail at specific CAPTCHA puzzles. Our work complements these early efforts by explicitly incorporating solving latency as a core metric, examining the reasoning traces of state-of-the-art MLLMs on diverse CAPTCHA types, and deriving concrete, defense-oriented design guidelines.

\section{Framework Development and Threat Model}
\label{sec:frameworkf}

This section describes a framework for evaluating CAPTCHA robustness against MLLMs, as shown in Figure~\ref{fig:framework}. We introduce the problem formulation, outline the attack strategy, and present the modeling assumptions used in this study.

\subsection{Threat Model and Attack Assumptions}
\label{subsec:formulation}
We consider a generic web service that deploys visual CAPTCHAs as part of its abuse-mitigation pipeline, for example, to protect account creation, content submission, or access to high-value resources. When a user reaches such a protected step, the service displays a CAPTCHA widget in the browser. Each CAPTCHA challenge has three pieces of information, which we will refer to throughout the paper: (i) $X$: one or more images shown in the CAPTCHA widget (a single picture, a grid of tiles, or a composed layout) that form the visual input to the solver; (ii) $q$: the human-readable instruction on the page (e.g., ``click all squares with traffic lights''); (iii) $y$: the unique correct answer stored and checked on the server.


In this setting, only the images $X$ and the instruction $q$ are visible in the browser; the ground-truth answer $y$ is never revealed to the client. In typical deployments, after the user submits an answer, the CAPTCHA widget returns a response token $r$ to the browser. The client then forwards $r$ to the relying party’s backend, which verifies $r$ with the CAPTCHA provider and uses the resulting pass/fail outcome to accept or reject the request.

As a result, benign users simply see the CAPTCHA, solve it by hand, and continue their normal interaction with the service. The malicious users (i.e., \emph{attacker}), in contrast, controls an automated client (or a script that drives a browser) with the goal of solving CAPTCHAs at scale so that large numbers of malicious requests can be automated. In this scenario, the web service together with its CAPTCHA deployment plays the role of the \emph{victim}.

We assume that the attacker can capture the images $X$ as rendered in browsers (e.g., via screenshots or image URLs), read human-facing instructions $q$, and submit answers on the client-side. However, the attacker cannot inspect/modify the server-side verification logic and never directly observes $y$. The attacker can leverage one or more MLLMs as automated CAPTCHA solvers. These models are accessed through black-box APIs, either from commercial providers or self-hosted open-source models. For each attempt, the attacker submits the images $X$ and a prompt derived from $q$, optionally enhanced with task-specific prompt optimizations such as clearer reasoning instructions or constrained output formats, and receives an instruction text response from the MLLMs. {We additionally consider an adaptive setting in which the attacker retains prior reasoning across attempts based on the accumulated interaction history together with binary pass/fail feedback.} The attacker has no access to gradients, internal activations, or model parameters; the models operate strictly as off-the-shelf oracles.

\subsection{MLLM Modeling Assumptions}
\label{sec:assumptions}

We evaluate MLLM CAPTCHA-solving capabilities on an offline dataset under four regimes representing plausible solver behaviors. Consistent with the formulation in Section~\ref{subsec:formulation}, the solver receives only the client-side inputs $(X, q)$ and produces a response $\hat{y}$. Results are determined by validating $\hat{y}$ against the ground truth $y$, without interacting with verification services. To systematically evaluate performance across multiple levels of adversary capability, we use the \texttt{Dice\_Count} task (where the requirement is to sum the pips on dice faces) and employ the following four specific strategies:

\textbf{(i) Direct Prompts (Exp1):} We forward the human-facing instruction $q$ directly to the model. For Dice\_Count, the prompt is: ``\textit{Sum up the numbers on all the dice.}'' This simulates a naive attacker who relies entirely on the MLLM's zero-shot instruction following without optimization.

\textbf{(ii) Optimized Prompts (Exp2):} The attacker utilizes a task-specific template that clarifies reasoning steps and enforces a machine-parseable format. For Dice\_Count, the prompt is expanded to: ``\textit{Identify all visible dice. Sum the pips on their top faces only. Output the final integer sum in JSON format (e.g., \{`answer': 14\}).}'' This isolates the model's visual reasoning capability from formatting errors.

\textbf{(iii) Until-correct Strategy (Exp3):} We model a determined attacker using an ``until-correct'' strategy with a budget of $k$ attempts. Upon failure, the solver requests a new instance of the same task type (e.g., a fresh Dice\_Coun puzzle) and retries using the \emph{Optimized Prompt} (as in Exp2). This accounts for the attacker's willingness to trade higher latency and cost for a successful bypass.

\textbf{(iv) Few-shot Prompts (Exp4):} To maximize solver performance on the hardest tasks, we prepend labeled examples to the optimized prompt. The model is presented with a context such as: ``\textit{User:  Sum the dice. Assistant: \{`answer': 5\}},'' followed by the target query.

To be noted here, the prompt template is fixed during evaluation to ensure consistent benchmarking for each model--task-type pair.
Our analysis relies on the following assumptions. First, conditional on a fixed model $m$, task type $t$ (defined in Section~\ref{subsec:dataset}), prompt template, and hyperparameters, we treat each attempt as an independent and identically distributed (i.i.d.) Bernoulli trial. This allows us to model the multi-attempt solving in Exp3 using standard binomial reasoning and to analytically estimate the probability of success within $k$ attempts from the single-shot results in Exp2.

Then, we focus strictly on the visual reasoning capabilities of the models. We do \emph{not} consider out-of-scope vectors such as compromising the web service, exploiting widget implementation bugs, or human-in-the-loop attacks. Finally, while emerging agentic frameworks allow for tool use (e.g., precise mouse control), the fundamental bottleneck remains the visual perception and reasoning capability of the underlying MLLM. If the model cannot correctly identify the target coordinates or counting logic, external tools cannot recover the correct solution. Therefore, our evaluation of the MLLMs can effectively bound the performance of agentic systems.

\section{Methods and Experiments}
\label{sec:methods}


\subsection{Dataset and Tasks}
\label{subsec:dataset}

{We conduct Exp1 to Exp4 on the recent CaptchaWorld dataset~\cite{captchaworld25}, which collects visual CAPTCHAs from both real-world services and synthetic environments. The raw dataset covers 20 task types and more than 400 instances; each instance contains the image(s), a natural-language instruction, and the ground-truth answer.
In this work, we apply cleaning and standardization to enable a unified MLLM-based evaluation: we remove two task types incompatible with our pipeline, correct and align labels and metadata, and normalize answer formats. The resulting main benchmark contains 18 task types and 378 instances.}

For the main benchmark, we group the 18 task types into four task families:
\begin{itemize}
    \item \textbf{Counting and aggregation}: Dice\_Count, Dart\_Count;
    \item \textbf{Pointing and path-based localization}: Place\_Dot, Geometry\_Click, Pick\_Area, Misleading\_Click, Click\_Order, Path\_Finder;
    \item \textbf{Grid selection and matching}: Bingo, Patch\_Select, Image\_Recognition, Select\_Animal, Unusual\_Detection, Object\_Match, Image\_Matching;
    \item \textbf{Relational and transformation puzzles}: Coordinates, Connect\_Icon, Rotation\_Match.
\end{itemize}

{Detailed task definitions, dataset statistics, and cleaning rules are deferred to Appendix~\ref{app:dataset-details}, with a summary of the supplemental set in Table~\ref{tab:dataset-overview-short}.} {We also consider a supplemental external evaluation set to test whether the main findings generalize beyond CaptchaWorld; details are provided later in Section~\ref{sec:external-adaptive}.}


\subsection{Evaluation Metrics}
\label{subsec:metrics}

We evaluate CAPTCHA solving along three key dimensions: accuracy, time cost, and finite-retry behavior. {Considering accuracy, we choose to use the first-time-pass-rate (Pass@1) as a metric. For example, Pass@1 measures the accuracy of the model when allowed strictly one attempt per instance, effectively representing the first-time pass rate. In addition, we further consider the end-to-end latency and finite-retry behavior as our metrics to understand MLLM latency and failure-retry performance in real-world situations.}

\textit{\textbf{Pass@1:}}
For each sample, let $\hat{y}_i$ be the answer produced by the model in its first attempt. We define
\begin{equation}
  \mathrm{Pass@1}
  \;=\;
  \frac{1}{N}\sum_{i=1}^{N}1\{\hat{y}_i = y_i\},
\end{equation}
where $N$ is the number of evaluation samples. We analogously compute Pass@1 per task type. Since each CAPTCHA instance has a single correct answer and the service returns only a binary pass/fail signal, Pass@1 aligns closely with how real deployments validate submissions. 

\textit{\textbf{End-to-end latency:}}
For each model call, we record the end-to-end latency from sending the request to receiving a complete, parseable answer. This metric directly captures the time between submitting a CAPTCHA to the model and learning whether it passed verification.

\textit{\textbf{Finite-retry behavior:}}
Real services typically cap the number of attempts per session. To summarize behaviour under a retry budget $k$, we report \emph{Success@k}, the probability that this attack succeeds within $k$ attempts. We derive these quantities from single-shot Pass@1 under a simple Bernoulli model described in Section~\ref{subsec:statlink}.

\subsection{Models and Inference Setup}
\label{subsec:models}

We evaluate seven representative MLLMs from major providers, covering both closed-source flagships and strong open-source vision--language models (Table~\ref{tab:eval-models}). For GPT-series models we consider different \texttt{reasoning\_effort} settings (e.g., \emph{Medium} vs.\ \emph{None}) to capture the effect of explicit reasoning modes.

To ensure comparability, we abstract a unified inference interface
\begin{equation}
  \mathrm{Invoke}(M, X, q; \theta) \rightarrow (a, \mathrm{meta}),
\end{equation}
where $M$ is the model, $X$ the CAPTCHA image(s), $q$ the instruction prompt, and $\theta$ shared decoding parameters (temperature, maximum output length). The returned answer $a$ is a short json text (e.g., ""target click position"":[{""x"":405,""y"":535}]); $\mathrm{meta}$ contains provider-specific metadata such as token counts and timestamps.

For each task type we constrain the output format to match the ground-truth answer (e.g., a single index, a coordinate pair, or a short integer). When APIs support JSON-mode or schema enforcement we use it, otherwise we instruct the model to output JSON-only and discard responses that contain extraneous text. For an error analysis purpose, we additionally ask the model to verbalize its reasoning in additional experiments. 

\begin{table}[t]
\caption{Multimodal models evaluated in our study.}
\begin{tabular}{lll}
\toprule
\textbf{Provider} & \textbf{Model} & \textbf{Snapshot} \\
\midrule
OpenAI      & GPT-5.1 (Medium)             & 2025-11-13 \\
OpenAI      & GPT-5.1 (None)               & 2025-11-13 \\
OpenAI      & GPT-5 (Medium)               & 2025-08-07 \\
Anthropic   & Claude Sonnet 4.5            & 2025-09-29 \\
Google      & Gemini-2.5-Pro               & 2025-01 \\
Google      & Gemini-2.5-Flash             & 2025-01 \\
Fireworks\textsuperscript{\dag}   & Qwen-3-VL-235B-Instruct      & 2025-09-24\\
\bottomrule
\end{tabular}
\footnotesize
\textsuperscript{\dag}Exp1--Exp4 used Fireworks. Exp5 used OpenRouter for the same Qwen3-VL-235B-A22B-Instruct model after Fireworks stopped serving it.

\label{tab:eval-models}
\end{table}

\begin{figure*}[t]
  \centering
  \includegraphics[width=1\linewidth]{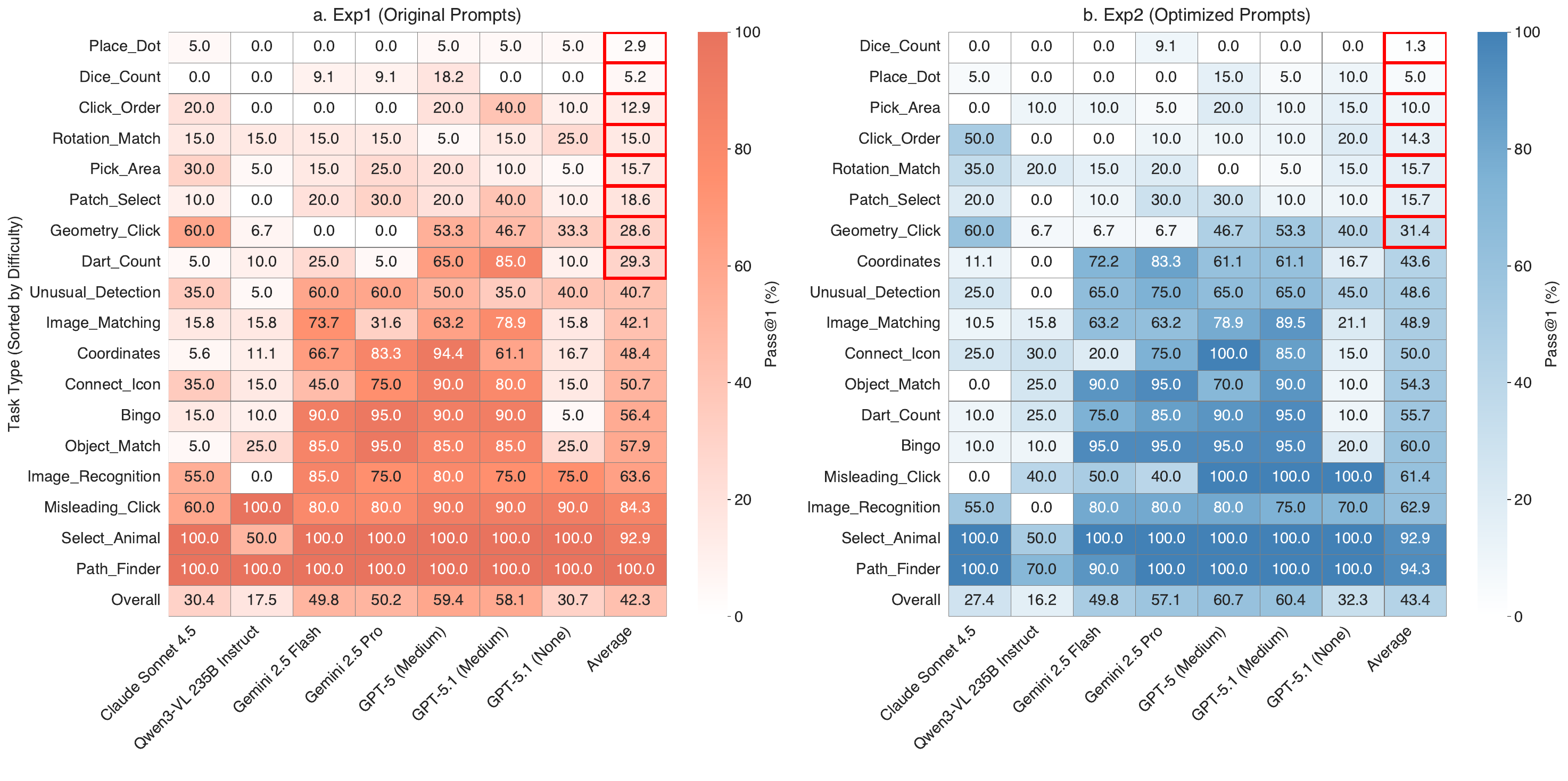}

  \caption{Heatmap of CAPTCHA task difficulty: (a) Exp1 (original prompts), and (b) Exp2 (optimized prompts). Task types (rows) are sorted by cross-model average Pass@1. Columns correspond to MLLMs, and each cell reports Pass@1 (\%).}

  \label{fig:heatmap-exp1-exp2}
\end{figure*}

\subsection{Experimental Protocol}
\label{subsec:pipeline}

Our evaluation consists of the following four complementary experiments. These four experiments (\textbf{Exp1} to \textbf{Exp4}) share the same dataset and metrics but differ in prompting strategies (defined in Section~\ref{sec:assumptions}) and retry settings.

\textit{\textbf{Exp1 (Direct Prompts):}}
We benchmark the raw capability of MLLMs by employing the \textit{Direct Prompts} strategy (Section~\ref{sec:assumptions}-i). For every model and task type, we perform a single invocation per sample using the original human-facing instruction. This establishes a baseline for off-the-shelf solver performance without prompt engineering.

\textit{\textbf{Exp2 (Optimized Prompts):}}
To isolate the effect of instruction tuning, we evaluate all models using the \textit{Optimized Prompts} strategy (Section~\ref{sec:assumptions}-ii). This experiment measures the performance gain achievable solely through clearer reasoning rules and standardized output formatting, serving as the primary basis for our accuracy analysis. {We also apply this same setup to the external evaluation set described in Section~\ref{sec:external-adaptive} to examine whether the insights observed on the main benchmark generalize beyond its original data sources.}

\textit{\textbf{Exp3 (Until-correct Strategy):}}
To assess operational viability, we model an attacker using an ``until-correct'' strategy with a retry budget $k$. Rather than performing redundant queries, we derive Success@$k$ and expected cost analytically from Exp2's single-shot accuracy, relying on the Bernoulli model detailed in Section~\ref{subsec:statlink}. We empirically validate the accuracy of this extrapolation on a subset of tasks.

\textit{\textbf{Exp4 (Few-shot Prompts):}}
For the specific task types identified as ``hard'' in Exp2, we apply the \textit{Few-shot Prompts} strategy (Section~\ref{sec:assumptions}-iii). By prepending two solved examples to the context, we test whether the observed failures are due to instruction misunderstanding or fundamental limitations in visual reasoning.


\textit{\textbf{Exp5 (Adaptive session-memory):}}
{We further evaluate whether session-level adaptation can improve the attacker's capability. In this setting, the attacker retains reasoning traces and submitted answers from earlier attempts within the same session, and may revise its local strategy after each failure. The only external feedback is the binary pass/fail outcome; ground-truth labels, target coordinates, counts, and corrective hints are never revealed. For each task type, we run independent memory-isolated rounds, resetting the conversation context between rounds. This experiment complements Exp3 by directly testing adaptive behavior rather than estimating independent fixed-prompt retries from single-shot Pass@1.  For Qwen3-VL in Exp5, we used the same Qwen3-VL-235B-A22B-Instruct model through OpenRouter, with a 2026-05-20 access date, because Fireworks no longer offered this model at this time.}

\subsection{Single-shot Accuracy and Finite-retry Cost}
\label{subsec:statlink}

\textbf{Non-adaptive finite-retry modeling (Exp3).} For a fixed model $m$ and CAPTCHA type $t$, all attempts use the same prompt template, decoding hyperparameters, and parsing rules, and are evaluated on instances drawn from the same dataset. Conversational memory and tool use are disabled, so randomness arises only from the sampled instance and the model’s internal sampling. Under these controlled conditions, it is natural to treat each attempt’s correctness as an independent Bernoulli trial with success probability $p_{m,t}$, which we estimate by the corresponding Pass@1 from Exp2.

Given this model, if an attacker is allowed at most $k$ attempts on type $t$, the probability that at least one attempt succeeds is
\[
\mathrm{Success@}k = 1 - (1 - p_{m,t})^{k},
\]
and the expected attempt count under this until-correct strategy is
\[
E[A] = \frac{1 - (1 - p_{m,t})^{k}}{p_{m,t}}.
\]
These quantities let us summarize, for each $(m,t)$, how a small retry budget amplifies single-shot accuracy and how many API calls an attacker is expected to make.

We estimate per-call monetary cost from provider price tables. Let $c_{\mathrm{in}}$ and $c_{\mathrm{out}}$ be the per-thousand-token prices for prompts and completions, and let $t_{i}^{\mathrm{in}}$ and $t_{i}^{\mathrm{out}}$ be the corresponding token counts for sample $i$. The cost of that invocation is
\[
\mathrm{cost}_{i}
=
\frac{t_{i}^{\mathrm{in}}}{1000} \, c_{\mathrm{in}}
+
\frac{t_{i}^{\mathrm{out}}}{1000} \, c_{\mathrm{out}}.
\]
Multiplying the average per-call cost by $E[A]$ yields an estimate of the expected cost per successful solve. For comparison, prior measurements of human CAPTCHA-solving services report roughly 80--90\% one-shot accuracy, solving times of a few to tens of seconds, and market prices on the order of 0.5--2 US cents per challenge~\cite{bursztein2010good,motoyama2010re,10646606}. We use these ranges as a reference when interpreting the cost--performance trade-offs of MLLM-based solvers.

{\textbf{Adaptive session-memory setting (Exp5).}} {The Bernoulli extrapolation above applies only to non-adaptive retries with a fixed per-attempt success rate. Exp5 instead directly evaluates whether an attacker can improve across attempts by using session memory together with binary pass/fail feedback. For each task type $t$, we run $n$ memory-isolated adaptive rounds. At the beginning of each round, the conversation context is reset. Within a round, the attacker is allowed up to $k$ attempts; after each failed attempt, it can review its previous reasoning traces, submitted answers, and the binary failure signal, then revise its prompt or local strategy before proceeding to another sampled instance from the same task type.}



{We report observed Success@k as the fraction of the $n$ rounds in which at least one of the first $k$ adaptive attempts succeeds:
\[
\widehat{\mathrm{Success@k}}_t
=
\frac{1}{n}\sum_{r=1}^{n}
\mathbf{1}\{\exists a \leq k:\hat{y}_{t,r,a}=y_{t,r,a}\},
\]
where $r$ indexes the memory isolated round, $a$ indexes the attempt within that round, and $y_{t,r,a}$ is the ground truth for the sampled instance. Because this metric is based on $k$ rounds of observations per task type, we use it as a stress test to assess whether feedback can qualitatively change the hardness conclusion under a limited retry budget, rather than as a precise deployment-wide success-rate estimate.}
\section{Experiment Results}
\label{sec:results}

The results reveal a clear difficulty gap across CAPTCHA types. Recognition-oriented and low-interaction tasks, such as object selection or simple grid matching, are already solved reliably by MLLMs with high accuracy, strong performance under small retry budgets, and low per-solve cost. In contrast, tasks requiring fine-grained localization, ordering, or counting remain challenging even for the strongest models and therefore serve as the most promising directions for defense-oriented deployment. We present our empirical findings in the order of RQ1–RQ3 and discuss defense implications (RQ4) in Section~\ref{sec:defense}.

\subsection{RQ1: Overall Solving Capability}

RQ1 asks how well MLLMs can solve diverse CAPTCHA tasks under realistic constraints such as limited time, retries, and cost. We first examine single-shot accuracy across models and task types, then connect these results to finite-retry success rates and time/cost trade-offs.

Figure~\ref{fig:heatmap-exp1-exp2}(a) summarizes Exp1, where models receive the original human-facing instructions. The cross-model average Pass@1 is $42.3\%$, showing that MLLMs already solve a substantial fraction of CAPTCHAs even with unoptimized prompts. Strong closed-source models (GPT-5, GPT-5.1, Gemini 2.5 Pro/Flash) reach around $50$–$60\%$ Pass@1, while weaker systems lag noticeably behind.

More importantly, there is a pronounced task-level hardness gap. A small set of types, including Place\_Dot, Dice\_Count, Click\_Order, Rotation\_Match, Pick\_Area, Patch\_Select, and Geometry\_Click, have cross-model averages well below $30\%$, with the first six mostly below $20\%$. In contrast, recognition-style tasks such as Path\_Finder and Select\_Animal already exceed $60\%$ on average, suggesting that they are easy targets for automated solvers.

\begin{figure}[t]
  \centering
  \includegraphics[width=\linewidth]{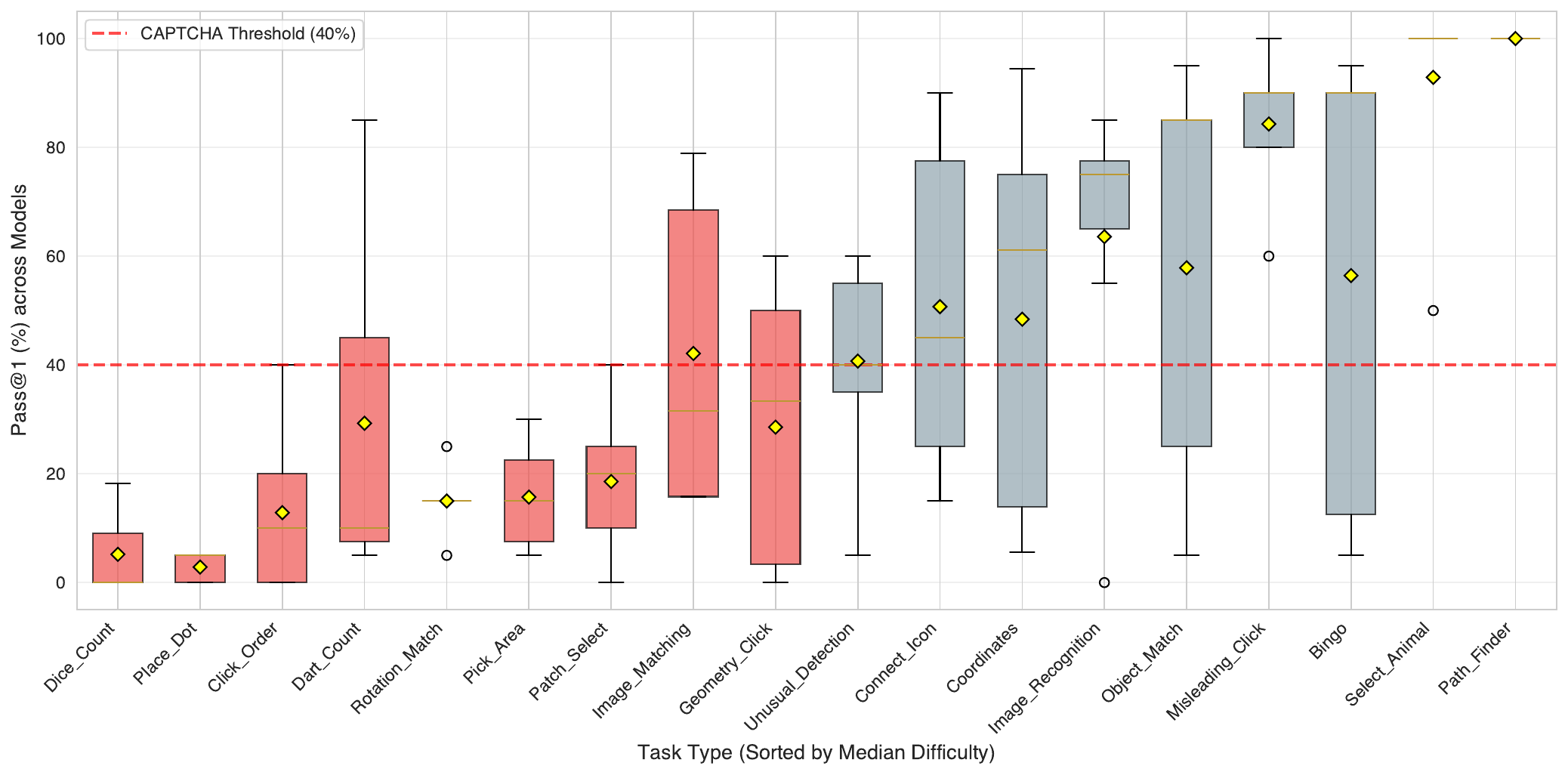}
  \caption{Cross-model Pass@1 distributions per task type in Exp1 (original prompts). Each box shows the spread across models. The dashed line marks a $40\%$ threshold.}
  \label{fig:stability-exp1}
\end{figure}

To verify that the hardness gap is not model-specific, we examine cross-model stability. Figure~\ref{fig:stability-exp1} plots Pass@1 distributions for each task type, with a dashed line at $40\%$ as a working CAPTCHA threshold. Tasks whose distributions lie well below this are consistently difficult across models, while those above it are reliably easy.

\begin{figure}[t]
  \centering
  \includegraphics[width=\linewidth]{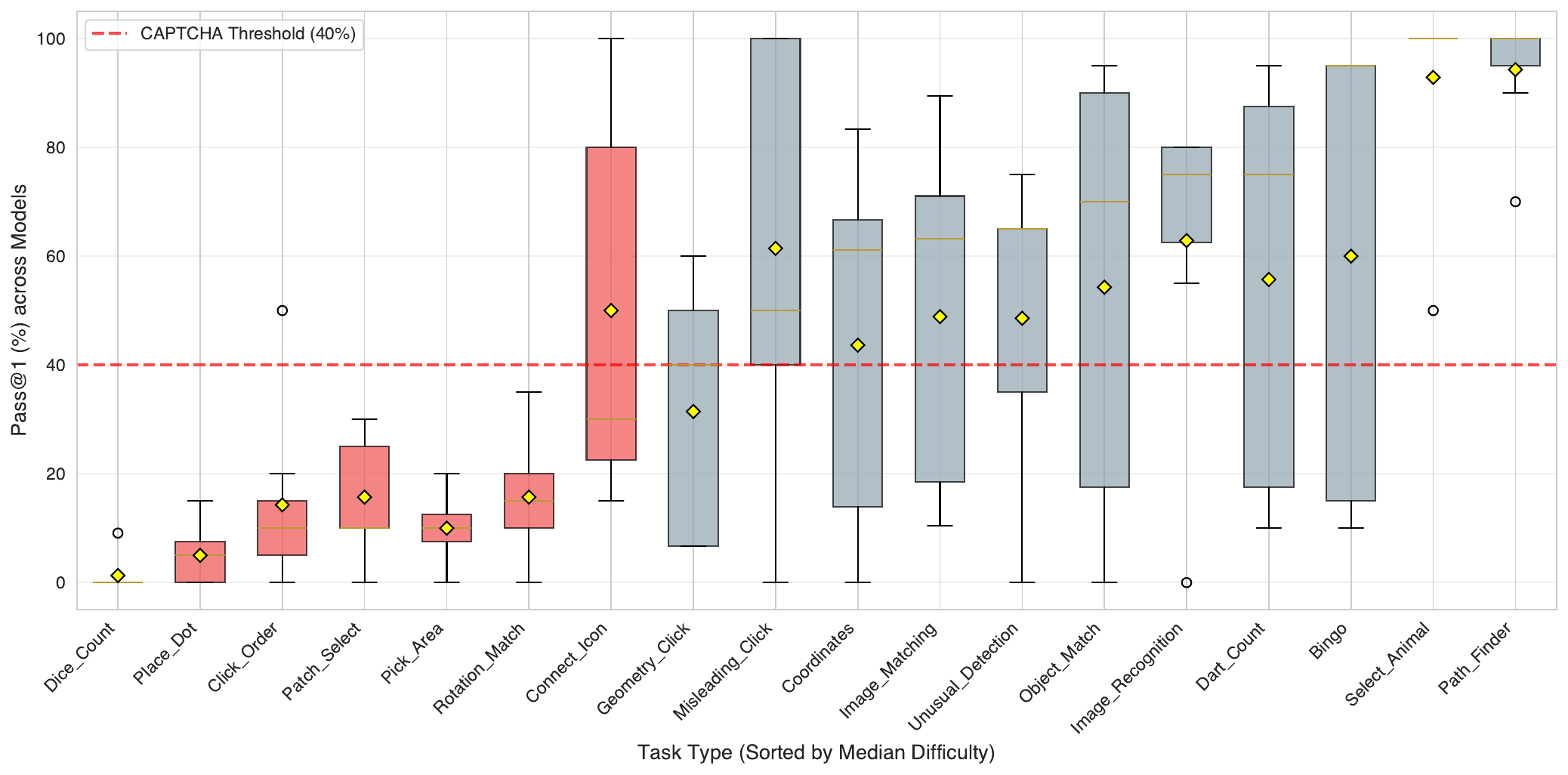}
  \caption{Cross-model Pass@1 distributions per task type in Exp2 (optimized prompts). Compared to Exp1, most recognition-style tasks rise well above the 40\% threshold, while the six hard task types remain low across models.}
  \label{fig:stability-exp2}
\end{figure}

Exp2 replaces the prompts with task-optimized instructions. As shown in Figures~\ref{fig:heatmap-exp1-exp2}(b) and~\ref{fig:stability-exp2}, prompt optimization yields modest gains for strong models (e.g., GPT-5 increases from $59.4\%$ to $60.7\%$, GPT-5.1 from $58.1\%$ to $60.4\%$), and slightly raises the overall average from $42.3\%$ to $43.4\%$. At the task level, however, the separation between easy and hard types becomes sharper: recognition-oriented tasks such as Bingo, Object\_Match, Image\_Recognition, and Path\_Finder move further above $60\%$, while six types—Dice\_Count, Place\_Dot, Pick\_Area, Click\_Order, Rotation\_Match, and Patch\_Select—remain below $20\%$ even after optimization. Geometry\_Click improves modestly but stays around $30\%$.

The experimental results from Exp1 and Exp2 reveal a distinct performance dichotomy. While prompt optimization further boosts the accuracy of recognition-based tasks, it fails to significantly improve performance on tasks requiring complex spatial grounding or counting. Consequently, the hardness gap observed in Exp1 is not a transient artifact of poor prompting but a stable characteristic of the tasks themselves. Based on this observation, we categorize the 18 task types into the following three security tiers to guide analysis:

\begin{itemize}
    \item \textbf{Broken types:} Task types whose cross-model average Pass@1 exceeds $40\%$ in both Exp1 and Exp2. This group contains recognition and simple grid-based families such as Path\_Finder, Select\_Animal, Misleading\_Click, Image\_Recognition, Bingo, Object\_Match, Unusual\_Detection, Image\_Matching, Coordinates, Connect\_Icon, and Dart\_Count.
    \item \textbf{Borderline types:} Geometry\_Click remains below $40\%$ on average but shows clear upward trends when prompts are optimized; we treat it as effectively broken.
    \item \textbf{Hard types:} Patch\_Select, Rotation\_Match, Click\_Order, Pick\_Area, Place\_Dot, and Dice\_Count stay below $20\%$ in both experiments with minimal improvement and are our candidate robust CAPTCHAs.
\end{itemize}

\begin{figure}[t]
  \centering
  \includegraphics[width=\linewidth]{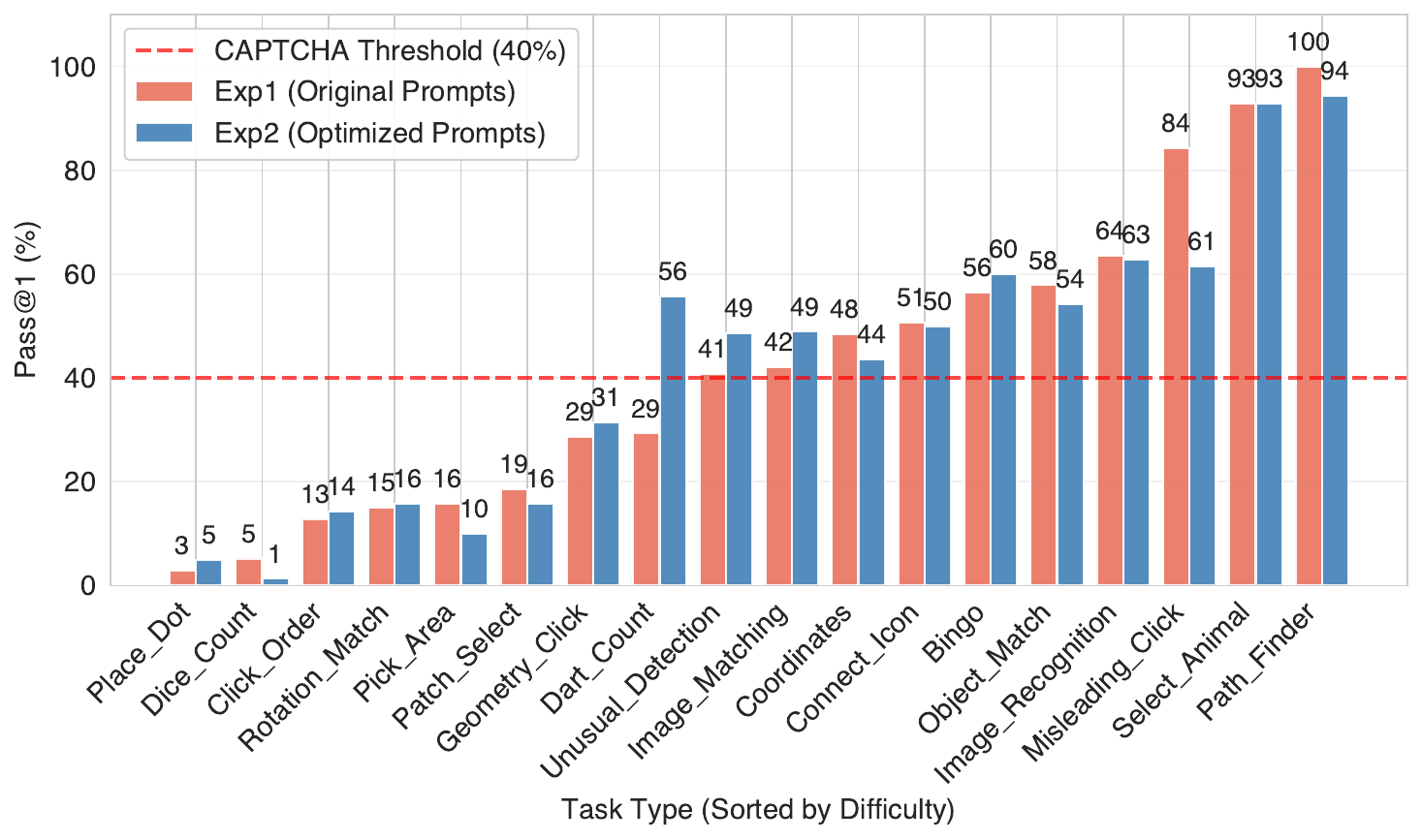}
  \caption{Per-task Pass@1 for GPT-5 (Medium) in Exp1 (original prompts) and Exp2 (optimized prompts). The dashed line at $40\%$ marks the hardness threshold.}
  \label{fig:comparison-gpt5}
\end{figure}

We focus on GPT-5 (Medium) as a representative strong solver, which attains the best overall performance. Figure~\ref{fig:comparison-gpt5} compares GPT-5’s Pass@1 under original and optimized prompts. Most recognition tasks lie well above the $40\%$ line (often above $80\%$) and benefit further from optimized prompts, whereas the six hard types remain below this threshold with only small or even negative changes. This shows that prompt engineering can polish already-vulnerable tasks but does not eliminate intrinsically hard ones.

\begin{figure}[t]
  \centering
  \includegraphics[width=\linewidth]{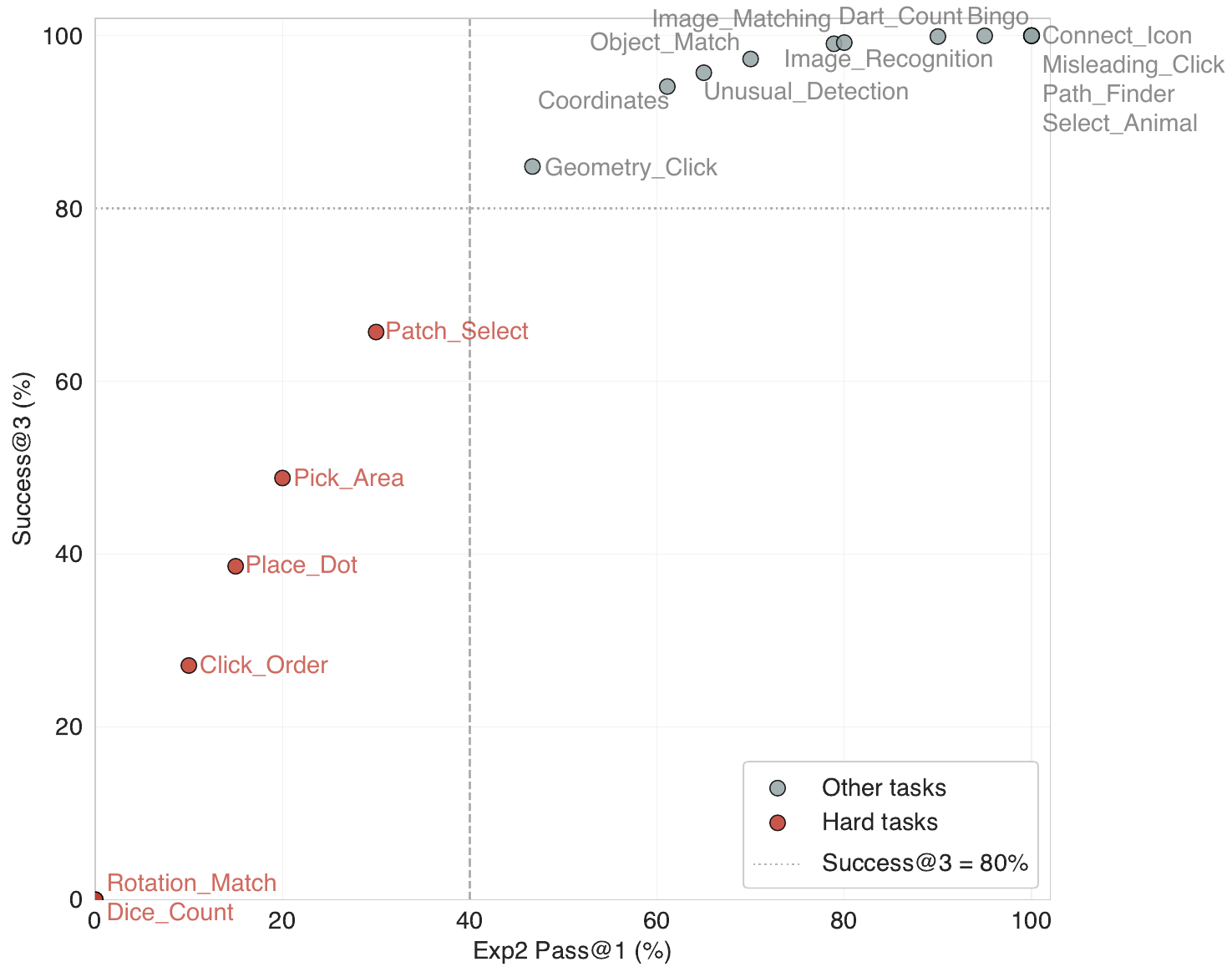}
    \caption{Mapping Exp2 Pass@1 to the finite-retry regime for GPT-5 (Medium). Each point is a task type; x-axis is single-shot Pass@1 and y-axis is the induced Success@3 under a three-attempt until-correct strategy.}
  \label{fig:exp3-mapping}
\end{figure}

To connect single-shot accuracy with realistic retry budgets, we map Exp2 Pass@1 to Success@3 using the statistical model in Section~\ref{subsec:statlink}. Figure~\ref{fig:exp3-mapping} shows that once Pass@1 exceeds $40\%$, Success@3 quickly rises above $80\%$, and tasks around $60\%$ Pass@1 already reach over $90\%$ Success@3. All broken types fall into this regime, implying that a small retry budget suffices to make them almost always solvable. The hard types remain in the lower-left corner: with Pass@1 below $30\%$, their Success@3 stays moderate even with three attempts, and Dice\_Count and Rotation\_Match remain effectively unsolvable.

\begin{figure}[t]
  \centering
  \includegraphics[width=\linewidth]{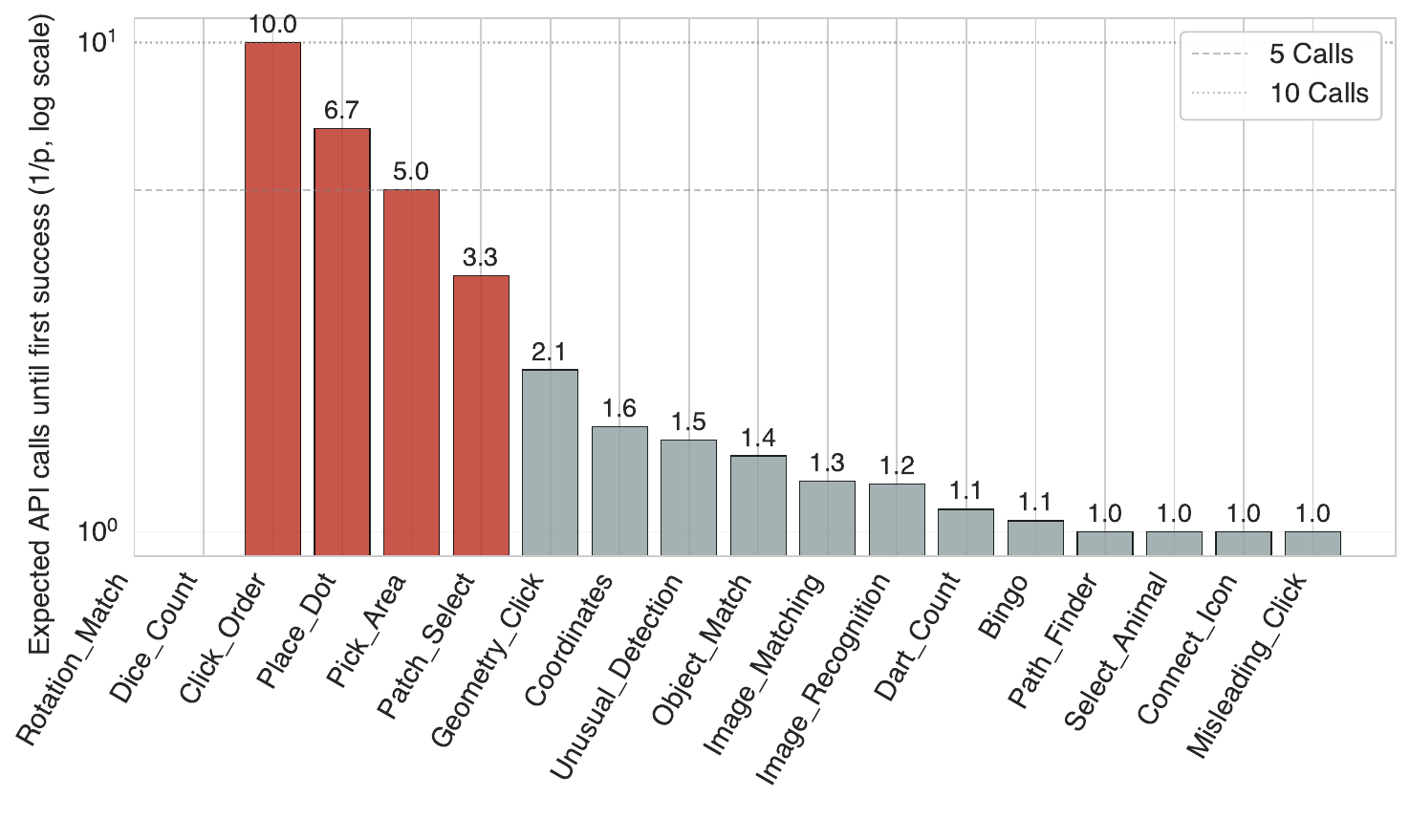}
  \caption{Expected number of API calls until the first success for GPT-5 (Medium), computed as $1/p$ based on Exp2 single-shot accuracy (log scale on the x-axis). Each point is a task type; reference lines at 5 and 10 calls highlight that hard tasks (shown in red in the plot) require significantly more calls than broken tasks.}
  \label{fig:exp3-expected-calls}
\end{figure}

Figure~\ref{fig:exp3-expected-calls} reports the expected number of API calls until the first success, again derived from Exp2. Broken types cluster between one and two calls on average, while the six hard tasks typically require between two and ten calls, with Dice\_Count and Rotation\_Match at the upper end. Thus, even without an explicit retry cap, reliably solving hard CAPTCHAs demands substantially more queries.

\begin{figure}[!htbp]
  \centering
  \begin{subfigure}[t]{0.84\linewidth}
    \centering
    \caption{Cost–performance frontier for GPT-5 in Exp1 and Exp2.}
    \includegraphics[width=\linewidth]{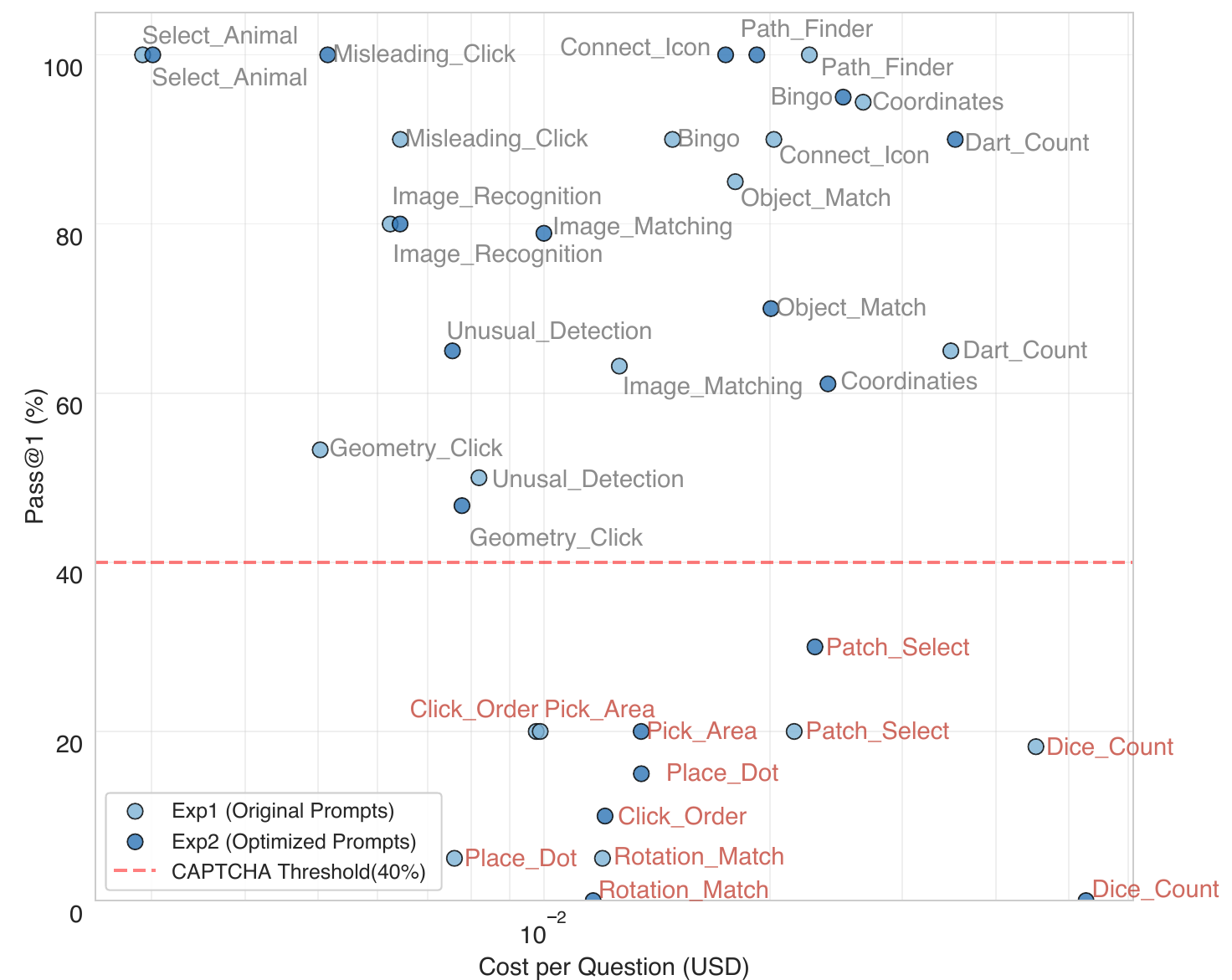}
    \label{fig:cost-frontier}
  \end{subfigure}

  \begin{subfigure}[t]{0.84\linewidth}
    \centering
    \caption{Time–performance trade-off for GPT-5 in Exp1 and Exp2.}
    \includegraphics[width=\linewidth]{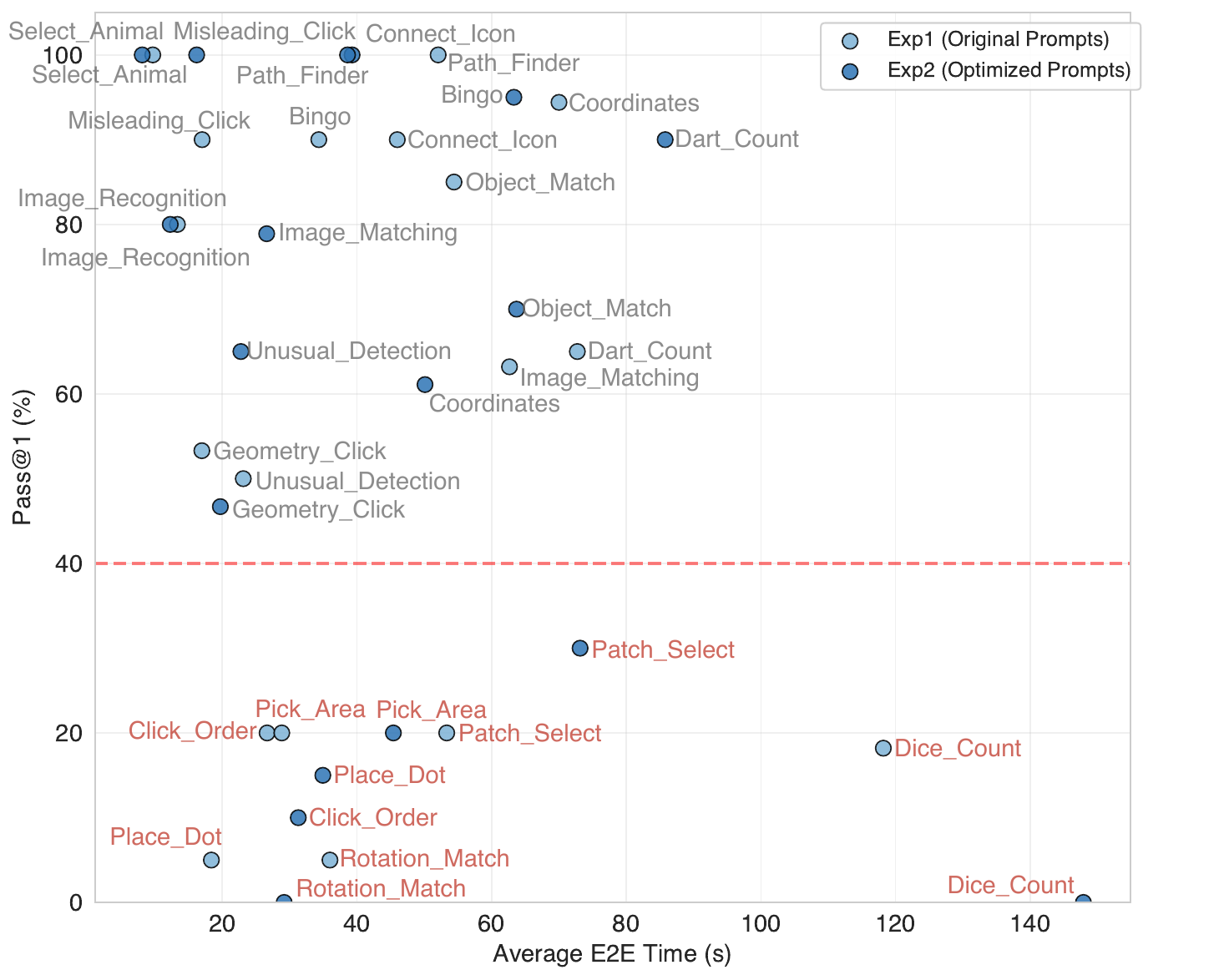}
    \label{fig:time-performance}
  \end{subfigure}

  \begin{subfigure}[t]{0.88\linewidth}
    \centering
    \hspace*{0.02\linewidth}%
    \caption{Expected cost per successful solve versus Pass@1 across CAPTCHA task types for GPT5}
    \includegraphics[width=\linewidth]{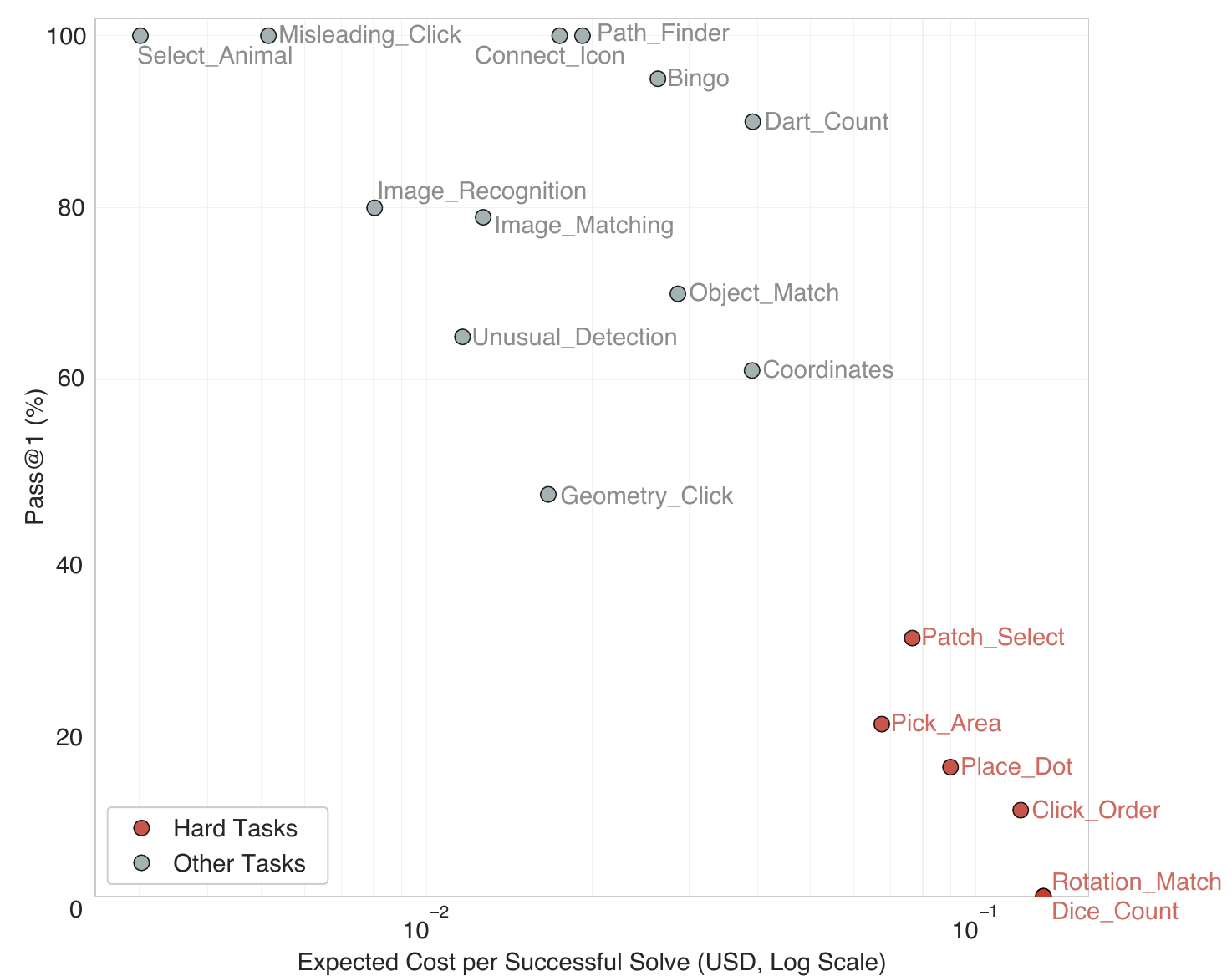}
    \label{fig:cost-success}
  \end{subfigure}
  \caption{Cost and latency trade-offs for GPT-5 across CAPTCHA task types.}
  \label{fig:cost-latency-overview}
\end{figure}

We incorporate monetary cost and latency. From an attacker’s perspective, RQ1 is not only about whether a CAPTCHA can be solved, but also whether solving it is economically and temporally viable at scale. Using the cost model in Section~\ref{subsec:statlink}, Figure~\ref{fig:cost-frontier} plots per-call cost versus Pass@1, while Figures~\ref{fig:time-performance} and~\ref{fig:cost-success} show E2E latency and expected cost per successful solve. Broken types occupy the favorable region: high Pass@1, E2E times on the order of tens of seconds, and expected costs per success less than 10 cents. The six hard types sit in the opposite regime: lower accuracy, similar or higher latency, and one to two orders of magnitude higher expected cost per successful solve. Under current pricing, they are therefore economically unattractive for large-scale automated solving.

As a result, Exp1–Exp3 and the cost/latency analysis answer RQ1: under realistic retry budgets and current API pricing, many recognition-style CAPTCHAs are already solvable cheaply and reliably, whereas the six hard task types remain both statistically and economically unattractive to automated MLLM-based solvers.

\subsection{RQ2: Impact of Prompting Strategies}
\label{rq2}

RQ2 asks how prompting strategies affect solver performance across CAPTCHA types. We therefore compare direct prompting (Exp1), task-optimized prompts (Exp2), and few-shot demonstrations (Exp4).

At the aggregate level, moving from original instructions to optimized prompts modestly improves overall Pass@1 for strong models (e.g., GPT-5 from $59.4\%$ to $60.7\%$ and Gemini 2.5 Pro from $50.2\%$ to $57.1\%$). Figures~\ref{fig:heatmap-exp1-exp2} and~\ref{fig:comparison-gpt5} show that these gains mainly push already-solvable recognition tasks further into the high-accuracy regime, while the six hard types remain well below the $40\%$ threshold. Prompt engineering thus amplifies capabilities where the model is already near the decision boundary but does not change which CAPTCHA families are intrinsically difficult.

We next ask whether few-shot demonstrations can substantially close the performance gap on the hard task types. In Exp4, we prepend two labeled examples per task type to the optimized prompt and evaluate GPT-5 under otherwise identical settings (Section~\ref{subsec:pipeline}). Table~\ref{tab:exp4-hard-tasks} summarizes the results.

\begin{table}[t]
\small
\centering
\caption{Few-shot (Exp4) performance of GPT-5 (Medium) on the six hard task types. E2E time is computed from the average per-question latency and reported in seconds.}
\begin{tabular}{lrr}
\toprule
\textbf{Task} & \textbf{Pass@1} & \textbf{Avg E2E (s)} \\
\midrule
Click\_Order      & 0.25 & 31.4 \\
Dice\_Count       & 0.00 & 92.2 \\
Patch\_Select     & 0.00 & 14.0 \\
Pick\_Area        & 0.00 & 38.7 \\
Place\_Dot        & 0.00 & 38.7 \\
Rotation\_Match   & 0.50 & 45.7 \\
\bottomrule
\end{tabular}
\label{tab:exp4-hard-tasks}
\end{table}

Few-shot guidance provides only limited benefit. Compared to Exp2, Click\_Order improves from $10\%$ to $25\%$ Pass@1 and Rotation\_Match from $0\%$ to $50\%$, but Dice\_Count, Patch\_Select, Pick\_Area, and Place\_Dot remain unsolved. At the same time, multi-turn prompts with embedded examples substantially increase latency: the average end-to-end time per query on these six tasks is about $43.4$\,s, with Dice\_Count exceeding $90$\,s. This corresponds to an average slowdown of roughly $2$–$3\times$ relative to Exp2. In other words, few-shot prompting does not reliably solve hard CAPTCHAs and makes each query considerably more expensive.

Taken together, the results of Exp1, Exp2 and Exp4 answer RQ2: task-specific prompt optimization and few-shot demonstrations improve MLLM solvers on already-vulnerable recognition CAPTCHAs, but do not qualitatively change which task families are hard. The same six hard types identified under direct prompting remain out of reach even with stronger prompting strategies.

\subsection{RQ3: Which CAPTCHAs Break, Which Survive, and Why?}

Combining the above results, we explain which CAPTCHA families current MLLMs reliably solve, which are close to being solved, and which remain comparatively robust.

\textit{\textbf{Broken and near-broken types:}}
Across Exp1 and Exp2, recognition-oriented and low-interaction tasks (e.g., object selection and coarse path finding) exhibit high Pass@1 for strong models and near-certain Success@3 with a small retry budget. These tasks require only coarse visual recognition with minimal spatial precision or multi-step reasoning, making them effectively broken under current MLLMs. Rotation\_Match is an intermediate case: it remains moderately difficult with only optimized prompts, but becomes largely solvable with two few-shot demonstrations in Exp4, reducing to simple pattern matching. We therefore treat it as a near-broken type that is likely to fall completely as model capabilities improve.

\textit{\textbf{Persistently hard types and shared structure:}}
After excluding broken and near-broken CAPTCHA types, five task types remain consistently hard across models and prompting regimes: Click\_Order, Place\_Dot, Pick\_Area, Dice\_Count, and Patch\_Select. Even GPT-5 with optimized prompts and few-shot examples exhibits low Pass@1, limited Success@3, and higher time and token consumption on these tasks. As discussed in Section~\ref{sec:defense}, they share several structural factors: a reliance on high-precision 2D localization, multi-object binding and ordering, and counting or aggregation under clutter. These properties stress aspects of spatial grounding and compositional reasoning that current MLLMs handle poorly.

\textit{\textbf{Recurrent error modes:}}
The reasoning traces logged for GPT-5 on these five task types exhibit recurring error modes:
\begin{itemize}
  \item \emph{Correct reasoning, incorrect discrete answer:} the model describes the right region or counting strategy, but the final coordinate, tile index, or numeric result does not match its own reasoning.
  \item \emph{Local correctness, global inconsistency:} the model correctly interprets local parts of the scene (e.g., icons or dice groups) but aggregates them inconsistently, leading to wrong click orders or off-by-one counts.
  \item \emph{Overconfident approximations in spatial decisions:} the model outputs locations that are ``approximately right’’ from a human perspective but fall outside the exact tolerance required by the verification logic.
\end{itemize}


In the following subsections, to understand the mechanisms behind these failures, we analyze representative reasoning traces on the hardest task types, such as the dice counting challenge or spatial localization.

\subsubsection{{Counting Challenge}}
The Dice\_Count task presents a $3\times2$ grid of panels, each containing several dice with pips and printed digits. The objective is to sum all visible top faces. Our analysis of Exp4 failures reveals that GPT-5 systematically overestimates the total by a wide margin, despite correctly articulating the task rules.
For example, in one representative failure, the model answers 92 while the ground truth is 69. Its reasoning trace exposes the root cause:
\begin{quote}
\textit{``I examined the six sub-images and counted only the top faces of each die \dots{} Subimage sums: top-left $= 5+4+1+5$, top-middle $= 2+2+4+3$, \dots{} bottom-right $= 3+5+7+7$. Adding these gives $15 + 11 + 17 + 11 + 16 + 22 = 92$.''}
\end{quote}
We analyzed four such sampled items, finding overestimations ranging from +9 to +26. The rationales consistently emphasize that only upward faces are counted and provide panel-wise decompositions that appear internally coherent. However, the model likely misinterprets side faces as top faces, double-counts dice that appear across panel boundaries, or misreads digits under perspective distortion. Once these local perceptual errors are propagated into the panel sums, the final aggregation step simply preserves the error. This suggests that the dominant difficulty is not the final arithmetic, but the robust perception and grouping of small objects in cluttered scenes.

\subsubsection{{Spatial Grounding Failures}}
Tasks requiring fine-grained spatial localization, such as Place\_Dot and Click\_Order, expose a fundamental disconnect between semantic understanding and coordinate prediction in MLLMs. We demonstrate the following three different tasks:

\textit{\textbf{Coarse Path Tracing in Place\_Dot.}}
\textit{Place\_Dot} requires placing a dot at the end of a car's path in a stylized 3D city grid. In several Exp4 failures, GPT-5 describes a detailed and correct procedure for tracing the path from the start pin to its terminal tip, yet selects a coordinate hundreds of pixels away from the target.
For instance, in one puzzle (target at $(290, 235)$), the model clicks $(565, 895)$---a Euclidean distance error of over 700 pixels. The model explains:

\begin{quote}
\textit{``I identified the bright green path connected to the car pin (start) at the top-right. Tracing the single continuous ribbon through its 90-degree turns leads to the opposite end near the lower central-left area. I clicked the visual center of that squared tip.''}
\end{quote}

Despite claiming to follow the path, the actual coordinates lie on a different segment or in a completely different region. This illustrates that while the global structure is understood conceptually, the model struggles to maintain a precise, metric-consistent representation of the polyline in image space, especially under perspective effects and 3D shading.

\textit{\textbf{Qualitative Area Estimation in Pick\_Area.}}
In Pick\_Area, the solver must identify the largest region among several enclosed shapes. GPT-5 frequently provides a plausible qualitative description of the partition structure but fails to perform precise geometric comparisons.
In one failure case, the correct largest region occupied a bounding box from $(200, 300)$ to $(510, 500)$. However, the model selected a point at $(660, 110)$ in a completely different, smaller cell, reasoning:

\begin{quote}
\textit{``The broadest region lies in the top-right... I clicked near the geometric center of this largest area.''}
\end{quote}
The reasoning traces indicate that GPT-5 approximates ``largest area'' based on coarse visual impressions of width and shape rather than strict geometric computation. When several regions have comparable sizes or complex boundaries, this qualitative heuristic leads to consistent mis-ranking.

\textit{\textbf{Boundary Violations in Click\_Order.}}
Similarly, Click\_Order requires clicking icons in a specific sequence. GPT-5 generally identifies the correct semantic order (e.g., light bulb $\to$ wheel $\to$ briefcase) but fails at the final discrete mapping. In one example, the model's predicted coordinates were visually close to the target icons but deviated by approximately 40--300 pixels, exceeding the tolerance radius. The model effectively ``solves the puzzle in words'' but fails to ground these words into acceptable pixel coordinates.

\subsubsection{{Contextual Hallucinations}}
One of the most surprising failure modes emerged in the Patch\_Select task, where the model is required to select grid patches containing a target object (e.g., a bridge). GPT-5 exhibited a recurring ``binding mistake'' where it confused the few-shot demonstration image provided in the prompt with the actual test image.

In an analysis of four distinct puzzles with different targets, e.g., bridge, hourglass, bench, maple leaf, GPT-5 consistently returned an empty set with high confidence. Strikingly, the ``raw observation'' segment of every trace described the exact same scene: a nighttime street festival with a large moon-like installation. This scene corresponded to the few-shot example, not the test query.
For example, when asked to find a \emph{bridge} in a new image, the model reasoned:
\begin{quote}
\textit{``I scan for classic bridge cues... The scene shows a nighttime street festival with a large illuminated sphere... but no continuous span with a walkable deck. Therefore, no grid cells contain a bridge.''}
\end{quote}
This task-level binding error arises from implicitly treating the example image as the query and is most prominent in Patch\_Select. It suggests that for dense grid tasks, strong few-shot guidance can paradoxically degrade performance by inducing contextual hallucinations, contributing directly to the near-zero success rate on this task type.

\subsection{{External Validation of the Core Findings}}
\label{sec:external-adaptive}

{The preceding results identify a stable boundary between CAPTCHA types that are already vulnerable to current MLLM-based solvers and those that remain more resistant. We next validate these core findings in three complementary ways: using a supplemental external evaluation set, using an adaptive session-memory test, and comparing our conclusions with prior solver and benchmark studies.}

{
\textit{\textbf{External dataset evaluation.}} To further examine whether the insights observed on this benchmark generalize beyond its original data sources, we additionally construct a disjoint supplemental external evaluation set of 80 instances from prior sources~\cite{liu2026nextgen,qi2026viperstrike}. We select these instances because they explicitly include the same failure-inducing properties identified in our main findings: counting under visual clutter, grid-level structural selection, and relation-based object-location binding. These categories therefore provide an external stress test of whether MLLM difficulty is driven by general structural properties rather than by the special data distribution in CaptchaWorld. The supplemental set follows the same instance schema and answer-format normalization as the main benchmark. Detailed task definitions, dataset statistics, and cleaning rules are deferred to Appendix~\ref{app:dataset-details}, with a summary of the supplemental set in Table~\ref{tab:dataset-overview-short}.
These three supplemental categories are revisited later in Table~\ref{tab:adaptive-hard-scope}, where we compare them against the six hard or near-hard task types from the main benchmark under the same GPT-5 adaptive setting.}

{
Table~\ref{tab:external-static-summary} summarizes the supplemental external evaluation set under the Exp2 setup. For each category, we report a pooled Pass@1 over all seven models and the best per-model Pass@1. Hole\_Counting has zero successful solves over 140 attempts. Symbol\_Count has 6 successes over 210 attempts, yielding an pooled Pass@1 of 2.9\%, with the best per-model reaching 10.0\%. Relation\_Match has 17 successes over 210 attempts, yielding an aggregate pooled Pass@1 of 8.1\%, with the best per-model reaching 16.7\%. All three external categories remain well below the 40\% threshold used in our main analysis. These results support the same core finding on externally sourced data: all three supplemental categories remain difficult for off-the-shelf MLLMs under optimized prompting, as their success rates remain well below the 40\% threshold.
}

\begin{table}[t]
\centering
\small
\caption{{Results on the external evaluation dataset. ``Pooled Pass@1'' aggregates attempts from all 7 MLLM-based model configurations within each category, ``Best per-model Pass@1'' reports the highest Pass@1 achieved by any single model.}}
\label{tab:external-static-summary}
\begin{tabular}{lrrrr}
\toprule
\textbf{Category} & \textbf{Pooled Pass@1} & \textbf{Best-model Pass@1} \\
\midrule
Hole\_Counting  & 0.0\% & 0.0\% \\
Relation\_Match & 8.1\% & 16.7\% \\
Symbol\_Count  & 2.9\% & 10.0\% \\
\bottomrule
\end{tabular}
\end{table}

{\textit{\textbf{Adaptive session-memory evaluation.}} Table~\ref{tab:adaptive-hard-scope} groups together the six hard or near-hard task types from the main CaptchaWorld benchmark and the three supplemental external categories introduced above. We test a GPT-5 (Medium) attacker that retains within-round history and updates its strategy using only binary pass/fail feedback.} {Table~\ref{tab:adaptive-hard-scope} compares observed Adaptive-Success@3 with the Bernoulli-Success@3 estimated from the same GPT-5 (Medium) Pass@1 reference. This is not a direct comparison, since Bernoulli estimate assumes independent non-adaptive retries, whereas adaptive results come from a five-round session-memory experiment. On average across the nine tasks, Success@3 increases only slightly, from 21.0\% under the Bernoulli to 24.4\% under adaptive evaluation. The main upward deviations occur on near-zero static tasks, while structurally hard tasks still remain low, including Hole\_Counting (0\%), Symbol\_Count (0\%), and Dice\_Count (20\%).}
{Thus, adaptive memory does not change the overall difficulty pattern: it can only help some boundary cases, but precise grounding, counting, and object-location binding remain much less reliable than simple recognition or selection tasks.}

\begin{table}[t]
\centering
\small
\caption{{GPT-5 (Medium) comparison of static and adaptive performance. The first six rows are hard or near-hard task types from the main CaptchaWorld benchmark, in the same order as Table~\ref{tab:exp4-hard-tasks}. The final three rows are supplemental external categories introduced in Table~\ref{tab:external-static-summary}.}}
\label{tab:adaptive-hard-scope}
\begin{tabular*}{\columnwidth}{@{\extracolsep{\fill}}lccc@{}}
\toprule
\textbf{Task} & $\mathbf{Pass@1}$ &
\begin{tabular}[c]{@{}c@{}}$\mathbf{Success@3}$\\ \textbf{(Bernoulli)}\end{tabular} &
\begin{tabular}[c]{@{}c@{}}$\widehat{\mathbf{Success@3}}$\\ \textbf{(Adaptive)}\end{tabular} \\
\midrule
Click\_Order    & 10\% & 27.1\% & 20\% \\
Dice\_Count     & 0\%  & 0.0\%  & 20\% \\
Patch\_Select   & 30\% & 65.7\% & 40\% \\
Pick\_Area      & 20\% & 48.8\% & 40\% \\
Place\_Dot      & 15\% & 38.6\% & 20\% \\
Rotation\_Match & 0\%  & 0.0\%  & 40\% \\
\midrule
Hole\_Counting  & 0\%  & 0.0\%  & 0\%  \\
Relation\_Match & 0\%  & 0.0\%  & 40\% \\
Symbol\_Count   & 3\%  & 8.7\%  & 0\%  \\
\bottomrule
\end{tabular*}
\end{table}

{\textit{\textbf{Comparison with external solvers and benchmarks.}}
Our study focuses on the native performance of off-the-shelf MLLMs in a controlled black-box setting, rather than on solver pipelines augmented with task-specific decomposition, fine-tuning, or additional engineering in prior works. We therefore compare our findings with recent CAPTCHA solver and benchmark studies~\cite{teoh2025captchas,deng2024oedipus,gao2021research,song2025reasoning,wu2025mca} only as contextual baselines, not as direct head-to-head measurements. In COGNITION, the solver interface is intentionally restricted to direct off-the-shelf MLLM API calls on raw CAPTCHA instances, without task decomposition, family specific modules, or richer agent orchestration.  In particular, systems such as Oedipus~\cite{deng2024oedipus} do not merely change the prompt; they alter the solver architecture by decomposing a CAPTCHA into structured intermediate substeps before invoking the LLM, thereby evaluating engineered attack capability rather than the native capability of an off-the-shelf model. Halligan et al.~\cite{teoh2025captchas}, for example, study an end-to-end agentic attacker in both closed world and live field settings, so the reported success rates reflect a richer interactive attack pipeline rather than the native one-shot behavior of a fixed off-the-shelf API. Earlier Visual Reasoning CAPTCHA~\cite{gao2021research} are often built around fixed CAPTCHA families or vendor templates, with modular recognition components and task engineering tailored to those structures rather than to a mixed benchmark of raw CAPTCHA instances. Benchmark suites such as MCA-Bench and CAPTCHA-X~\cite{wu2025mca,song2025reasoning} broaden task coverage and are valuable for identifying which CAPTCHA families remain difficult, but they do not provide a matched protocol for measuring direct off-the-shelf API performance under a uniform black-box interface. Table~\ref{tab:external-sota-baselines} summarizes these differences and Appendix~\ref{app:external-baselines} provides a fuller methodological comparison. As a result, these comparisons support the external validity of our main conclusion: recognition and simple selection tasks are increasingly easy for modern solvers, whereas tasks requiring precise grounding, counting, or relation binding remain more resistant.}

\section{{Defense and Limitation Discussion}}
\label{sec:defense}

{Based on our experimental evaluation, we identify visual CAPTCHA task types that remain challenging for state-of-the-art MLLMs, even under optimized prompting, retries, and few-shot examples. We then address RQ4: how can web service providers keep CAPTCHA schemes effective against increasingly capable MLLM-based solvers? To answer this question, we distill the key structural \emph{factors} underlying the robust task types, translate them into a practical defense methodology, and provide a preliminary validation showing how this methodology can strengthen a vulnerable CAPTCHA task.}

\subsection{Hardness Factors from Robust Task Types}


{To understand why certain CAPTCHA task types remain resistant to MLLMs, we examine the structural properties. As shown in Section~\ref{sec:results}, while recognition-based tasks are often solvable, a smaller subset continues to resist automated solvers. From the observations, we distill three key factors underlying robustness, which expose a persistent gap between semantic reasoning and visual grounding in current models.}

\textit{\textbf{Fine-grained spatial localization:}}
Across the five robust task types, the solver must localize targets on a continuous canvas rather than select from a few discrete options. The answer can be a specific point (Click\_Order and Place\_Dot) or a region defined by geometric or semantic cues (Pick\_Area). Verification logic allows a tolerance window, but the model must still map language to precise coordinates. In practice, MLLMs often describe the correct region yet output points or areas that fall outside the acceptable zone.

\textit{\textbf{Counting and aggregation beyond pattern recognition:}}
Dic\-e\_Count and counting-based variants of Pick\_Area couple visual recognition with explicit counting and light-weight arithmetic. Models frequently recognize local objects but miscount or mis-aggregate them, or make small arithmetic slips in the final answer. But, humans are reliable at counting small sets and performing simple sums, making these tasks structurally more robust than pure recognition or binary decisions.

\textit{\textbf{Stability across models and prompting regimes:}}
Click\_Order, Place\_Dot, Pick\_Area, Dice\_Count, and Patch\_Select remain hard for all evaluated models under direct prompting and optimized prompts, and they stay hard even with task-specific few-shot demonstrations. In contrast, many recognition-style CAPTCHAs are highly sensitive to prompt wording and quickly become solvable under modest prompt engineering. Robust types thus appear constrained by deeper perceptual and reasoning limitations rather than by prompt-level misunderstandings, making them better candidates for long-lived deployment.

\noindent{These hardness factors provide the basis for a practical hardening methodology: rather than relying on generic distortion, operators can redesign CAPTCHA tasks to selectively increase demands on grounding, counting, and multi-step consistency.}


\subsection{{A Practical Defense Methodology}}
\label{subsec:guidelines}

{Drawing on the hardness factors above, we recommend a simple four-step defense methodology for practitioners. \textbf{First}, replace coarse recognition or discrete choice mechanics with tasks that require fine-grained visual grounding. \textbf{Second}, when localization alone is insufficient, further increase difficulty by coupling perception with counting, aggregation, or other lightweight multi-step reasoning. \textbf{Third}, when stronger protection is needed, compose multiple hardness factors within a single challenge so that the solver must complete several dependent subtasks correctly. \textbf{Fourth}, validate that the resulting task remains usable for legitimate users, monitor solver performance after deployment, and periodically refresh templates or instructions as model capabilities evolve. The four steps below elaborate this workflow.}

{\textit{\textbf{First, favor fine-grained localization over discrete choice:}}
Operators can replace discrete-choice mechanics with tasks that require precise coordinate prediction, thereby shifting the challenge from coarse recognition to fine-grained visual grounding. This transformation directly targets one of the most persistent weaknesses of current MLLM-based solvers. Unlike discrete selection, e.g., choosing a tile from a grid, continuous-space tasks require the user to pinpoint a specific location within a strict tolerance window. As evidenced by the robustness of Place\_Dot and Click\_Order, forcing the model to ground semantic instructions into precise pixel coordinates introduces a significantly higher error rate than standard multiple-choice formats.}

{\textit{\textbf{Second, couple perception with counting and simple arithmetic:}} When localization alone is insufficient, operators can further increase difficulty by coupling perception with counting or lightweight aggregation, thereby requiring the solver not only to recognize visual elements but also to correctly quantify or combine them.
Instead of asking which tiles contain a certain object, prefer tasks that require counting small objects or aggregating visual evidence into a numeric answer, as in Dice\_Count or counting-based Pick\_Area. Even simple arithmetic, e.g., summing small numbers, interacts poorly with current visual pipelines while being easy for humans, but counts should remain small to avoid excessive user burden.}

{\textit{\textbf{Third, combine multiple hardness factors in a single challenge:}} Operators can compose multiple hardness factors within the same challenge so that the solver must complete several dependent sub-tasks correctly.
Robustness increases when a single challenge incorporates multiple difficulty factors. A single challenge can require users to locate specific features and verify a condition simultaneously. For example, interact with objects in a prescribed order, and optionally report an aggregate quantity. Such compositions forces the MLLM to perform a sequence of dependent sub-tasks consisting of attribute recognition followed by precise localization where any single error leads to verification failure.}

{\textit{\textbf{Finally, preserve usability and plan for evolution:}} Throughout this process, operators should validate that the hardened task remains usable for humans, monitor solver performance after deployment, and periodically refresh templates or instructions as model capabilities evolve.
Static templates eventually become vulnerable to fine-tuning. Operators can implement dynamic rotation strategies to maintain hardness. This includes regularly varying rendering styles to disrupt model overfitting and periodically rephrasing instructions, such as changing ``\textit{Select the largest region" to "Identify the area with the maximum surface}.'' This can prevent attackers from stabilizing their system prompts. Furthermore, operators should actively monitor solving metrics. A sudden spike in success rates or a convergence of solving times can indicate that an automated solver has bypassed the current scheme, necessitating an immediate template update.}

{Taken together, these four steps provide a practical defense methodology: increase grounding precision, add lightweight counting or aggregation when needed, compose multiple hardness factors for stronger protection, and iteratively validate the resulting design against both security and usability objectives. By integrating multiple hardness factors, such as asking a user to ``\textit{click on all visible eyes of the cat},'' we combine counting with fine-grained localization. This composition forces the solver to execute a chain of dependent reasoning steps, where a single failure in  logic or grounding invalidates the attempt. This significantly raises the technical and economic bar for attackers by rendering cheap, off-the-shelf MLLMs ineffective and necessitating the development of specialized, expensive solvers. In the following section, we provide a preliminary validation of these principles by hardening a representative vulnerable task.}

\subsection{{Preliminary Validation for Defense}}
\label{subsec:validation}
{To illustrate how this methodology can be applied in practice, we instantiate it on the Select\_Animal task type, which was classified as ``Broken'' in our baseline evaluation.
In our baseline evaluation (Exp1 and Exp2), this task with state-of-the-art MLLMs achieved near-perfect Pass@1 rates (e.g., GPT-5 $>95\%$). The original task involved selecting grid tiles containing a target animal, which MLLMs solved easily via coarse semantic recognition.
Following the hardening methodology in Section~\ref{subsec:guidelines}, we transformed this task by replacing coarse grid selection with fine-grained localization and by introducing an implicit counting requirement.}

\begin{itemize}
    \item \textit{Fine-grained spatial localization:} Instead of selecting a tile, the solver is required to ``click on all \textit{visible} eyes'' of the target animal. This shifts the requirement from discrete classification to continuous-space coordinate prediction with a tight tolerance ($15$ px).
    \item \textit{Implicit counting and existence detection:} The target animals are presented in varying poses (e.g., profile vs.\ frontal). The solver must visually ground the number of visible eyes (1 or 2) rather than relying on the prior knowledge that animals typically have two eyes.
\end{itemize}

{We evaluated this hardened variant against the same suite of MLLMs using task-optimized prompts (Exp2). The results, summarized in Table~\ref{tab:defense-validation}, show a dramatic collapse in solver performance. In the following analysis, we dissect these results to verify the security gain (\textit{Effectiveness}), confirm the defense mechanism through trace analysis (Failure Modes), and ensure that this added security does not come at the cost of user experience (\textit{Usability Implications}).}

\begin{table}[t]
\small
\centering
\caption{Defense validation on the hardened Select\_Animal task. The proposed design changes reduced the success rate of SOTA models from near-perfect (Baseline) to near-zero (Hardened), effectively mitigating the threat.}
\label{tab:defense-validation}
\begin{tabular}{lrr}
\toprule
\textbf{Model} & \textbf{Baseline Pass@1} & \textbf{Hardened Pass@1} \\
\midrule
GPT-5 (Medium)      & $>95\%$ & \textbf{0.0\%} \\
GPT-5.1 (Medium)    & $>95\%$ & \textbf{0.0\%} \\
GPT-5.1 (None)      & $>95\%$ & \textbf{0.0\%} \\
Gemini 2.5 Pro      & $>90\%$ & \textbf{0.0\%} \\
Claude Sonnet 4.5   & $>90\%$ & 23.3\% \\
Qwen3-VL          & $>80\%$ & 20.0\% \\
\bottomrule
\end{tabular}
\end{table}

\textit{\textbf{Effectiveness:}}
As shown in Table~\ref{tab:defense-validation}, the success rates for GPT-5, GPT-5.1 (both Medium and None), and Gemini 2.5 Pro dropped to \textbf{0.0\%}. Notably, the failure of GPT-5.1 (None) confirms that the difficulty is structural and cannot be bypassed simply by reducing reasoning effort or relying on faster inference modes. Even the strongest performer on this new task, Claude Sonnet 4.5, achieved only $23.3\%$, falling well below the $40\%$ threshold for a ``Broken'' task. 

\textit{\textbf{Failure Analysis:}}
Analysis of the reasoning traces reveals that models frequently succumbed to two error modes predicted by our guidelines: (1) \emph{Hallucination based on priors}, where models predicted the location of a second eye on a profile face where only one was visible, e.g., clicking empty space on the animal's cheek. (2) \emph{Imprecise grounding}, where clicks landed on the face but missed the specific tolerance window of the eye.

\textit{\textbf{Usability Implications:}}
Crucially, this increase in security does not impose a proportional burden on human users. Identifying and pointing to visible eyes is a natural perceptual task for humans. While it requires slightly more precision than simple grid selection, it avoids the visual strain induced by adversarial noise, trading image degradation for a light-weight cognitive task that humans intuitively excel at. This validation confirms that by structurally aligning CAPTCHA mechanics with MLLM weaknesses—specifically imprecise localization and hallucination—we can restore security guarantees without degrading the human user experience.

{While demonstrated on Select\_Animal, these results validate the broader efficacy of the structural guidelines proposed in Section~\ref{subsec:guidelines}. The complete failure (\textbf{0.0\%}) of both GPT and Gemini series, combined with the suppressed performance of Claude Sonnet 4.5 ($23.3\%$) and Qwen3-VL ($20.0\%$), confirms that the defense is robust across diverse model families. This cross-model collapse indicates that coupling fine-grained localization with implicit counting successfully exploits the shared neuro-symbolic gap in current MLLMs. By retrofitting such primitives onto existing tasks, operators can force attackers to abandon scalable, general-purpose APIs in favor of costly, specialized solutions, fundamentally altering the economic asymmetry of the attack.}

\subsection{Limitations}


{Although \textit{Open CaptchaWorld} provides a broad and useful benchmark for evaluating the security of visual CAPTCHA systems against MLLMs, it does not fully capture the long tail of edge-case, highly customized, or domain-specific CAPTCHA implementations that may appear in practice. Our findings should therefore be interpreted as evidence on a representative benchmark of modern visual CAPTCHA designs rather than as a comprehensive evaluation of all real-world deployments; they also reflect the current capability landscape of MLLMs, which may shift as models continue to improve, and do not exhaust the space of stronger custom or agentic attackers. In particular, the current benchmark does not yet include direct comparison against engineered solvers such as Oedipus or Halligan, which limits the scope of our empirical comparison to off-the-shelf MLLM settings; we plan to extend the benchmark in future work to support such comparisons. \textit{Open CaptchaWorld} combines curated prior sources with our unified benchmark construction and evaluation pipeline, but the resulting benchmark remains limited in scale. {The adaptive session-memory experiment should be interpreted as a stress test rather than a deployment-wide probability estimate, and both model capability and attack cost may shift over time.} Notably, while the dataset is sufficient to reveal clear empirical trends, the quantitative results should be interpreted within the scope of this benchmark and the evaluated model/pricing snapshot, rather than as precise universal estimates of attack success rates or costs for every CAPTCHA implementation in the wild.}

\section{Conclusion}
\label{sec:conclusion}
{This paper examines how modern MLLMs challenge the security of visual CAPTCHAs, evaluating seven representative models across 21 real-world task types under a realistic black-box threat model that accounts for retries, latency, and cost. Our results reveal a sharp hardness gap: most recognition oriented CAPTCHAs are now reliably solvable, often at a low cost and within a few retries, while a compact set of tasks requiring precise localization, cross-panel consistency and counting remains consistently robust even under optimized prompting, few-shot demonstrations or adaptive session-memory. Analysis of model reasoning traces shows that failures on these hard tasks stem from persistent weaknesses in spatial grounding and object–position binding, which informs a set of defense guidelines that emphasize continuous space localization, perception combined with counting, and the inclusion of multiple difficulty factors within a single challenge. We validate these strategies by hardening a representative vulnerable task, demonstrating that structural defenses can reduce MLLM success rates from near-perfect levels (>95\%) to zero. Crucially, compared to traditional defenses that rely on aggressive image blurring and cause visual fatigue, our redesign maintains high visual clarity. While it introduces a slight shift towards fine-grained perceptual interaction, it leverages natural human capabilities rather than frustrating users with degraded images.}

\appendix
\section*{Ethical Considerations}
This research studies the security posture of visual CAPTCHA systems against emerging MLLMs. While our work demonstrates methods to bypass existing security mechanisms, we have taken strict measures to ensure our methodology adheres to ethical research principles and minimizes potential harm.

\textit{\textbf{Interaction with Online Services:}}
A primary ethical concern in CAPTCHA research is the potential impact on the availability and integrity of live web services. To mitigate this risk, our evaluation was primarily conducted on the \textit{Open CaptchaWorld} dataset, which aggregates historical and synthetic CAPTCHA instances.
We did not perform high-volume, automated attacks against production registration flows or protected endpoints of active web services. The black-box attack setting described in our methodology simulates an attacker's perspective using offline instances, ensuring no disruption to real-world server infrastructure or degradation of user experience for legitimate users.

\textit{\textbf{Usage of MLLM APIs:}}
Our experiments utilized commercial APIs (e.g., OpenAI, Google, Anthropic). We adhered to the standard usage tiers and rate limits provided by these services. While automated interaction with web interfaces is often restricted, our research focuses on the safety evaluation of the models' reasoning capabilities rather than exploiting the service providers themselves. All API usage was paid and authenticated, ensuring fair compensation for the computational resources consumed.

\textit{\textbf{Human Subjects:}}
This study implies comparisons with human performance (e.g., human-like cost and latency ). We explicitly state that no new human subject experiments were conducted for this paper. All baselines regarding human solving accuracy, speed, and cost were derived from established prior literature and publicly available market data from CAPTCHA-solving services. Therefore, Institutional Review Board (IRB) approval was not required.

\textbf{Stakeholder Impacts.}
Beyond direct research risks, our results have implications for several
stakeholders. Legitimate users may benefit from stronger protection
against automated abuse, but harder CAPTCHA designs can also increase
completion time, error rates, and accessibility burden. In particular,
increasing task granularity to avoid MLLM solutions should not be
interpreted as a recommendation to make challenges arbitrarily more
complex. Deployments should preserve visual clarity, keep counting and
interaction sequences small, provide accessible alternatives, and measure human completion time and failure rates before rollout. Researchers may benefit from reproducible evaluation artifacts for follow-up studies, but such artifacts also create dual-use risk; accordingly, our release focuses on offline evaluation and omits production-oriented automation or service-specific bypass code. Companies operating CAPTCHA-protected services can use our findings to guide redesign and monitoring, but should treat the proposed hardening strategies as a security--usability trade-off that requires staged deployment and periodic reassessment.

\textit{\textbf{Responsible Disclosure:}}
Consistent with minimizing harm, we initiated disclosure to affected stakeholders upon identifying clear weaknesses and shared non-weaponized reproduction materials for verification. {Specifically, we disclosed the findings to Google, OpenAI, Anthropic, hCaptcha, Cloudflare, and Alibaba. For Google, we submitted an AI VRP report, which had been reviewed and closed by Jan.~26,~2026; Google indicated that the submission was considered out of scope rather than a technical vulnerability in Google's infrastructure. For OpenAI, we sent a disclosure message in Jan.~29, 2026 summarizing the study setting, the main findings, and high-level mitigation directions, and OpenAI replied with guidance on the appropriate reporting route. For Anthropic, we sent a similar disclosure in Jan.~28, 2026, and Anthropic acknowledged receipt. For hCaptcha, we sent a disclosure in early February 2026, and hCaptcha replied that the report had been forwarded to the relevant internal team. For Cloudflare, we sent a disclosure message in Jan.~29, 2026 describing the findings, the offline evaluation setting, and possible mitigations, and we received a routing or intake response. For Alibaba, we likewise sent a disclosure message in Jan.~30, 2026 with the same scope and level of detail, and we received a routing or intake response. Accordingly, we believe that publication is justified because these findings help defenders better understand the current limitations of visual CAPTCHAs and support the development of more robust defenses.} 

\textit{\textbf{Dual-Use and Code Release:}}
To balance reproducibility with safety, our released code serves as a research framework for evaluating model capabilities, lacking the engineering optimizations required for weaponized, large-scale bot operations. We believe that the defensive insights specifically the design guidelines for MLLM-resistant CAPTCHAs outweigh the risks. While the MLLM capabilities we evaluate are already available to the public, documenting their impact on CAPTCHAs is essential for the security community to develop countermeasures. 

\section*{Open Science}
The public artifact necessary to evaluate the contributions of this paper is permanently available through Zenodo at \url{https://doi.org/10.5281/zenodo.20406852}. The released package includes the offline CAPTCHA/MLLM evaluation framework, the cleaned benchmark data and supplemental external categories used in the paper, recorded results and outputs for reproducing figures and tables, and supporting reproducibility infrastructure such as tests, dependency files, and artifact-building utilities. Consistent with our open science and safety commitments, the release is limited to evaluation and verification artifacts. It excludes local secrets, provider credentials, live-service automation, and production-oriented CAPTCHA bypass infrastructure; these excluded materials are not required to evaluate the paper's claims.

\bibliographystyle{plainurl}
\bibliography{references}

\appendix

\section{Model API Pricing}
\label{app:api-pricing}

Table~\ref{tab:api-pricing} summarizes the token-based prices used in our cost analysis. 
All values are in USD per 1{,}000 tokens and correspond to the public pricing at the time we ran our experiments.
We only account for prompt and completion tokens reported by the providers; internal ``reasoning'' or hidden tokens (when applicable) are not observable and are therefore excluded from our estimates.

\begin{table}[b]
  \small
  \centering  
  \caption{Token-based model pricing (USD per 1{,}000 tokens).}
  \begin{tabular}{llrr}
    \toprule
    \textbf{Model (API name)} & \textbf{Input} & \textbf{Output} \\
    \midrule
    gpt-5 & 0.00125 & 0.01000 \\
    gpt-5.1 & 0.00125 & 0.01000 \\
    \midrule
    claude-sonnet-4-5 & 0.00300 & 0.01500 \\
    \midrule
    gemini-2.5-pro & 0.00125 & 0.01000 \\
    gemini-2.5-flash & 0.00030 & 0.00250 \\
    \midrule
    qwen3-vl-235b-instruct & 0.00022 & 0.00088 \\
    \bottomrule
  \end{tabular}
  \label{tab:api-pricing}
\end{table}


\section{Dataset Description}
\label{app:dataset-details}

{Our benchmark contains 458 visual CAPTCHA instances across 21 task types (see Table~\ref{tab:dataset-overview-short}). Each instance consists of one or more images (PNG/JPG) and a structured \texttt{ground\_truth.json} entry with an English prompt, a short description, and machine-interpretable labels (click coordinates, grid indices, option indices, numeric counts, or rotation angles). Click-based tasks use pixel coordinates with a small tolerance radius; grid-based tasks use zero-based row-major indices; multi-answer tasks are stored as sets or sequences. In addition to cleaned CaptchaWorld benchmark, we construct a supplemental external evaluation set for further validation. These instances follow the same basic schema and standardization principles as the main benchmark, including metadata alignment, provenance and license checks, and answer-format normalization, but are kept separate from the benchmark. }



\begin{table*}[htbp]
  \centering
  \caption{{Task types and brief descriptions in our dataset, including supplemental external evaluation categories.}}
  \footnotesize
  \begin{tabularx}{\textwidth}{@{}l c X@{}}
    \toprule
    Task type & \# inst. & Brief description \\
    \midrule
    Bingo               & 25 & Swap two cells in a $3\times 3$ emoji grid to complete a line of identical symbols. \\
    Click\_Order        & 10 & Click a set of icons in the same order as shown in a separate reference image. \\
    Connect\_icon       & 20 & Choose which arrow connection matches the dotted-line pattern in the reference diagram. \\
    Coordinates         & 18 & Select the option where a character sits at the seat indicated in the reference image. \\
    Dart\_Count         & 20 & Choose the dartboard whose darts add up to a given target number. \\
    Dice\_Count         & 11 & Read a composite dice scene and output the total number of visible pips. \\
    Geometry\_Click     & 15 & Click the specified geometric element (e.g., a letter relative to a shape). \\
    Image\_Matching     & 19 & Match an object in the left image to the correct choice on the right. \\
    Image\_Recognition  & 20 & In a $3\times 3$ grid, select all images containing a specified object class. \\
    Misleading\_Click   & 10 & Click to continue while avoiding prominently highlighted ``do not click'' regions. \\
    Object\_Match       & 20 & Adjust a discrete object count until it matches the reference image. \\
    Patch\_Select       & 10 & In a $5\times 5$ patch grid, select all patches containing a target object. \\
    Path\_Finder        & 10 & Choose the path that moves an object to a cross-marked destination. \\
    Pick\_Area          & 30 & Click on the largest outlined region in a complex scene. \\
    Place\_Dot          & 32 & Place a dot at the end of a car's path along a drawn trajectory. \\
    Rotation\_Match     & 48 & Rotate an object so that its orientation matches a reference direction. \\
    Select\_Animal      & 30 & In a small grid, select all cells containing the target animal (e.g., fox). \\
    Unusual\_Detection  & 30 & Select images where an animal has a mismatched head and body. \\
    \addlinespace
    Symbol\_Count~\cite{liu2026nextgen}
                         & 30 & Count specified symbols in an occluded repeating grid. \\
    Relation\_Match~\cite{qi2026viperstrike}
                         & 30 & Choose the object crop matching a relational prompt in a 3D scene. \\
    Hole\_Counting~\cite{liu2026nextgen}
                         & 20 & Select grid tiles whose shapes contain the requested number of holes. \\

    \bottomrule
  \end{tabularx}
  \label{tab:dataset-overview-short}
\end{table*}

\section{Comparison with External Studies}
\label{app:external-baselines}
{Table~\ref{tab:external-sota-baselines} provides contextual comparison between COGNITION and recent CAPTCHA solvers and benchmarks. Beyond the high-level performance trends summarized in the main text, these studies also align with several of our task-level observations. Halligan reports that continuous actions and 3D spatial reasoning remain weaker than discrete selection and swap-style tasks~\cite{teoh2025captchas}, which is consistent with our failures on \textit{Place\_Dot} and \textit{Click\_Order}, where models often describe the correct semantic target but fail to output verifier-valid coordinates. Oedipus and earlier specialized visual-reasoning solvers further suggest that semantic difficulty alone is not a reliable defense once a CAPTCHA can be decomposed into stable, task-specific subproblems~\cite{deng2024oedipus,gao2021research}. This supports our design emphasis on combining semantic understanding with precise grounding, object-location binding, counting, and verifier-sensitive output constraints, rather than relying only on high-level reasoning complexity. Finally, larger benchmarks such as MCA-Bench and CAPTCHA-X show that interaction depth, exact spatial grounding, and multi-step structure continue to separate easier and harder CAPTCHA types~\cite{wu2025mca,song2025reasoning}, reinforcing the broader applicability of the structural factors identified in our benchmark.}

\begin{table*}[htbp]
\centering
\footnotesize
\caption{{Contextual comparison with state-of-the-art CAPTCHA solvers and larger benchmarks. These are not direct head-to-head comparisons to COGNITION because several prior systems evaluate engineered solver pipelines, rather than the native capability of off-the-shelf MLLM APIs in our black-box setting. For example, Oedipus decomposes CAPTCHA solving into structured intermediate substeps before invoking the LLM, whereas COGNITION measures direct off-the-shelf model performance.}}
\label{tab:external-sota-baselines}
\begin{tabularx}{\textwidth}{
  @{}
  >{\raggedright\arraybackslash}p{0.17\textwidth}
  >{\raggedright\arraybackslash}p{0.20\textwidth}
  >{\raggedright\arraybackslash}X
  >{\raggedright\arraybackslash}X
  @{}
}
\toprule
\textbf{System / study} &
\textbf{Solver type} &
\textbf{Reported setting} &
\textbf{Reported result} \\
\midrule

\textbf{This Work}
&
Direct Off-the-shelf MLLM APIs
&
18-task benchmark plus provenance-checked external set; static/adaptive analyses
&
Best $\sim$60\% Pass@1; external categories $<$17\%; adaptive Success@3 $<$40\%  \\
\addlinespace
\textbf{Halligan}~\cite{teoh2025captchas}
&
Agentic VLM solver
&
2,600 closed-world challenges; 3,000 live field tasks
&
60.7\% closed-world solve rate; 70.6\% field solve rate \\

\addlinespace
\textbf{Oedipus}~\cite{deng2024oedipus}
&
DSL-guided decomposition pipeline over an LLM
&
Reasoning CAPTCHA benchmark with decomposed substeps
&
63.5\% reported average success rate \\

\addlinespace
\textbf{Earlier Visual Reasoning CAPTCHA}~\cite{gao2021research}
&
Task-specific modular pipelines
&
Fixed visual-reasoning or vendor-specific CAPTCHA families
&
High success on target families; limited mixed-benchmark coverage \\

\addlinespace
\textbf{MCA-Bench / CAPTCHA-X}~\cite{wu2025mca,song2025reasoning}
&
Benchmark suites and agentic VLM attackers
&
Multimodal CAPTCHA and high-difficulty spatial CAPTCHA benchmarks
&
Simple visual tasks highly solvable; spatial and multi-step tasks harder \\

\bottomrule
\end{tabularx}
\end{table*}

\end{document}